\nonstopmode
\documentclass[]{pasj02} 
\usepackage[switch,mathlines]{lineno}
\makeatletter
\providecommand{\@LN@col}[1]{}
\makeatother

\usepackage{natbib} 
\usepackage{lscape}
\usepackage{url}
\usepackage{rotating}
\usepackage{graphicx} 
\usepackage{ulem}
\usepackage{xcolor}
\jyear{2026}
\Received{2026/08/21}
\Accepted{}

\begin{document} 

\title{
Diversity of Ionized Gas Structures in Nearby Metal-poor Dwarf Galaxies
}

\author{
 Yuki Takagishi,\altaffilmark{1}
 Takuya Hashimoto,\altaffilmark{1,2,3}
 Kazuya Matsubayashi, \altaffilmark{4}
 Matthew Hayes, \altaffilmark{5}
 Aida Wofford,\altaffilmark{6}
 Masato Hagimoto, \altaffilmark{7}
 Akio K. Inoue, \altaffilmark{8,9}
 Ken Mawatari, \altaffilmark{8,9}
 Yurina Nakazato,\altaffilmark{10}
 Wataru Osone,\altaffilmark{1}
 Yuma Sugahara, \altaffilmark{8,9}
 Yoshiki Toba, \altaffilmark{11,12,13}
 Hidenobu Yajima, \altaffilmark{1,14}
 Shunsuke Honda,\altaffilmark{1,3}
 Hiroshi Matsuo,\altaffilmark{15,16}
 and 
 Naoki Yoshida\altaffilmark{17,18,19}
 }
\altaffiltext{1}{
Division of Physics, Faculty of Pure and Applied Sciences, University of Tsukuba, Tsukuba, Ibaraki 305-8571, Japan
}
\altaffiltext{2}{
Tsukuba Institute for Advanced Research (TIAR), University
of Tsukuba, 1-1-1 Tennodai, Tsukuba, Ibaraki, 305-8577, Japan
}
\altaffiltext{3}{
Tomonaga Center for the History of the Universe (TCHoU), Faculty of Pure and Applied Sciences, University of Tsukuba, Tsukuba, Ibaraki 305-8571, Japan
}
\altaffiltext{4}{
Institute of Astronomy, School of Science, The University of Tokyo,
2-21-1 Osawa, Mitaka, Tokyo 181-0015, Japan
}
\altaffiltext{5}{
Stockholm University, Department of Astronomy and Oskar Klein Centre for Cosmoparticle Physics, AlbaNova University Centre, SE-10691, Stockholm, Sweden
}
\altaffiltext{6}{
Universidad Nacional Aut\'onoma de M\'exico, Instituto de Astronom\'ia, A.P. 106, 22800 Ensenada, B.C., M\'exico
}
\altaffiltext{7}{
Department of Physics, Graduate School of Science, Nagoya University, Nagoya, Aichi 464-8602, Japan
}
\altaffiltext{8}{
Waseda Research Institute for Science and Engineering, Faculty of Science and Engineering, Waseda University,3-4-1 Okubo, Shinjuku, Tokyo 169-8555, Japan
}
\altaffiltext{9}{
Department of Pure and Applied Physics, School of Advanced Science and Engineering, Faculty of Science and Engineering, Waseda University, 3-4-1 Okubo, Shinjuku, Tokyo 169-8555, Japan
}
\altaffiltext{10}{
Center for Computational Astrophysics, Flatiron Institute, 162 5th Avenue, New York, NY 10010, USA
}
\altaffiltext{11}{
Department of Physical Sciences, Ritsumeikan University, 1-1-1 Noji-higashi, Kusatsu, Shiga 525-8577, Japan
}
\altaffiltext{12}{
Academia Sinica Institute of Astronomy and Astrophysics, 11F of Astronomy-Mathematics Building, AS/NTU, No.1, Section 4, Roosevelt Road, Taipei 10617, Taiwan
}
\altaffiltext{13}{
Research Center for Space and Cosmic Evolution, Ehime University, 2-5 Bunkyo-cho, Matsuyama, Ehime 790-8577, Japan
}
\altaffiltext{14}{
Center for Computational Sciences, University of Tsukuba, 1-1-1 Tennodai, Tsukuba, Ibaraki 305-8577, Japan
}
\altaffiltext{15}{
National Astronomical Observatory of Japan, 2-21-1 Osawa, Mitaka, Tokyo 181-8588, Japan
}
\altaffiltext{16}{
The Graduate University for Advanced Studies (SOKENDAI), 2-21-1 Osawa, Mitaka, Tokyo 181-8588, Japan
}
\altaffiltext{17}{
Department of Physics, The University of Tokyo, 7-3-1 Hongo, Bunkyo, Tokyo 113-0033, Japan
}
\altaffiltext{18}{
Kavli Institute for the Physics and Mathematics of the Universe (WPI), UT Institute for Advanced Study, The University of Tokyo, Kashiwa, Chiba 277-8583, Japan
}
\altaffiltext{19}{
Research Center for the Early Universe, School of Science, The University of Tokyo, 7-3-1 Hongo, Bunkyo, Tokyo 113-0033, Japan
}

\KeyWords{galaxies: dwarf -- galaxies: ISM -- galaxies: starburst}

\maketitle
\begin{abstract}
We investigate whether optical and far-infrared [O\,{\sc iii}] emission from nearby metal-poor dwarf galaxies can be represented by a homogeneous one-zone ionized-gas model with a single electron temperature and density. Our sample comprises five galaxies from the \textit{Herschel} Dwarf Galaxy Survey: HS~1222+3741, SBS~0335--052E, POX~186, Haro~11, and I~Zw~18. We combine galaxy-integrated or nearly galaxy-integrated [O\,{\sc iii}] $\lambda4363$ and $\lambda5007$ measurements from Seimei/KOOLS-IFU observations and published or archival spectroscopy with \textit{Herschel}/PACS [O\,{\sc iii}] 88~$\micron$ measurements. Because [O\,{\sc iii}] $\lambda4363$ is not detected in HS~1222+3741, the quantitative analysis is based on the remaining four galaxies. SBS~0335--052E and Haro~11 lie near or slightly beyond the low-density boundary of the one-zone diagnostic. Their nominal line ratios favor effective densities of $n_{\rm e}<1~{\rm cm}^{-3}$, while conservative treatment of the uncertainties allows values up to approximately $40$ and $10~{\rm cm}^{-3}$, respectively. These remain substantially below densities inferred from independent diagnostics. By contrast, POX~186 and I~Zw~18 show no significant discrepancy between the optical--far-infrared [O\,{\sc iii}] and low-ionization optical diagnostics. Additional optical and ultraviolet diagnostics show that inferred densities can span several orders of magnitude within a galaxy. Representative two-zone models reproduce the [O\,{\sc iii}] $\lambda4363$, $\lambda5007$, and 88~$\micron$ emission in SBS~0335--052E and Haro~11 by combining relatively dense gas with cooler, low-density gas. The low-density component contributes approximately 61\% and 72\% of the 88~$\micron$ luminosity, but only 14\% and 23\% of the $\lambda5007$ luminosity, respectively. These solutions are not unique and may represent a broader unresolved distribution of gas conditions. Our results show that temperatures and densities inferred from integrated one-zone analyses are effective quantities and that similar diagnostic discrepancies can arise in nearby metal-poor galaxies.
\end{abstract}

\section{Introduction}
\label{sec:intro}

Understanding the physical conditions of ionized gas in the interstellar medium (ISM) is essential for studies of galaxy formation and evolution. Key quantities include the electron temperature, $T_{\rm e}$, electron density, $n_{\rm e}$, gas-phase metallicity, and dust attenuation, which characterize the thermal and chemical state of the ionized ISM and are closely linked to star formation and chemical enrichment \citep[e.g.,][]{Kewley2019}. Emission lines from ionized gas provide powerful diagnostics of these conditions.

Rest-frame optical and ultraviolet emission lines are widely used to characterize ionized gas in both nearby and distant galaxies. The [O\,{\sc iii}] $\lambda4363/\lambda5007$ ratio constrains $T_{\rm e}$ and forms the basis of the direct-$T_{\rm e}$ metallicity method, while density-sensitive ratios such as [O\,{\sc ii}] $\lambda3729/\lambda3726$, [S\,{\sc ii}] $\lambda6716/\lambda6731$, [Ar\,{\sc iv}] $\lambda4711/\lambda4740$, and [C\,{\sc iii}] $\lambda1907$/C\,{\sc iii}] $\lambda1909$ constrain $n_{\rm e}$ over different density regimes \citep[e.g.,][]{Dinerstein1985,Osterbrock2006,Kewley2019}. With JWST, these diagnostics are now accessible even for galaxies in the epoch of reionization \citep[e.g.,][]{Curti2023,Nakajima2023,Isobe2023,Morishita2024,Sanders2026,CrespoGomez2025}.

Far-infrared fine-structure lines provide a complementary probe of the ionized ISM. Density-sensitive ratios such as [O\,{\sc iii}] 52/88~$\micron$ and [N\,{\sc ii}] 122/205~$\micron$ constrain $n_{\rm e}$ \citep[e.g.,][]{Kewley2019,Yang2020,Nakazato2023,Decarli2025}. Owing to their low excitation energies and long wavelengths, these lines are less sensitive to $T_{\rm e}$ and dust attenuation than optical collisionally excited lines. Such diagnostics have long been applied to Galactic H\,{\sc ii} regions and nearby galaxies with ISO and \textit{Herschel} \citep[e.g.,][]{Brauher2008,Cormier2015,Diaz-Santos2017,Herrera-Camus2018,Spinoglio2022}, while ALMA has extended them into the epoch of reionization \citep[e.g.,][]{Sugahara2021,Sugahara2022,Killi2023,Harikane2025}. JWST and ALMA now enable direct comparisons of optical and far-infrared diagnostics across cosmic time.

A particularly useful diagnostic combines the optical [O\,{\sc iii}] $\lambda4363$ and $\lambda5007$ lines with the far-infrared [O\,{\sc iii}] 88~$\micron$ line. Because all three transitions arise from the same ionic species, O$^{++}$, their ratios probe the temperature and density of the O$^{++}$-emitting gas without dependence on elemental abundance ratios. The [O\,{\sc iii}] $\lambda4363/\lambda5007$ ratio primarily constrains $T_{\rm e}$, while [O\,{\sc iii}] 88~$\micron/\lambda5007$ depends on both $T_{\rm e}$ and $n_{\rm e}$. The critical density of [O\,{\sc iii}] 88~$\micron$, $\sim5\times10^{2}~{\rm cm}^{-3}$ at $T_{\rm e}\sim10^{4}$~K, is much lower than that of [O\,{\sc iii}] $\lambda5007$, $\sim10^{6}~{\rm cm}^{-3}$ \citep[e.g.,][]{Osterbrock2006}. Under a homogeneous one-zone assumption, both ratios should be reproduced by a single $T_{\rm e}$ and $n_{\rm e}$. Positions near or beyond the low-density boundary of the model therefore provide a sensitive test of whether a single set of physical conditions adequately represents the integrated [O\,{\sc iii}] emission. This diagnostic has been applied to nearby H\,{\sc ii} regions and galaxies \citep[e.g.,][]{Chen2023,Chen2026} and to high-redshift galaxies observed with JWST and ALMA \citep[e.g.,][]{Stiavelli2023,Harshan2024,Fujimoto2024,Usui2025,Harikane2025,Takechi2026}.

Galaxy-integrated spectra can mix ionized gas spanning different temperatures, densities, ionization conditions, and attenuation. In several high-redshift galaxies, the observed [O\,{\sc iii}] 88~$\micron$ emission is stronger than expected from the physical conditions inferred from the optical [O\,{\sc iii}] lines under a one-zone assumption. \citet{Usui2025} showed that the [O\,{\sc iii}] $\lambda4363$, $\lambda5007$, and 88~$\micron$ ratios of COS-2987 at $z=6.81$ cannot be reproduced by a single $T_{\rm e}$ and $n_{\rm e}$, even after correcting for dust attenuation. Their two-component model reproduced the elevated [O\,{\sc iii}] 88~$\micron/\lambda5007$ ratio, with compact, hot, dense gas dominating the optical lines and more extended, cooler, low-density gas contributing strongly to the far-infrared emission. Similarly, \citet{Harikane2025} found that electron densities inferred from optical [O\,{\sc ii}] doublets in galaxies at $z\simeq6$--7 can exceed those inferred from the [O\,{\sc iii}] 52~$\micron$/88~$\micron$ ratio, suggesting that optical and far-infrared diagnostics may preferentially trace different density regimes.

Such differences need not imply sharply separated physical components. Recent observations of the Orion Nebula, H\,{\sc ii} regions, and nearby star-forming galaxies reveal a systematic hierarchy among electron-density diagnostics that can be reproduced by broad underlying density distributions \citep{MendezDelgado2026}. In this picture, individual line ratios preferentially weight different parts of the density distribution rather than measuring a unique representative electron density, even for diagnostics involving the same ionic species.

Whether similarly complex density structure is common in nearby metal-poor galaxies that share key properties with high-redshift star-forming systems remains unclear. \citet{Harikane2025} found broad agreement between optical- and far-infrared-based $n_{\rm e}$ for most nearby galaxies in their comparison sample, with the Seyfert galaxy NGC~1068 as a notable exception. The optical densities were derived from galaxy-integrated [S\,{\sc ii}] $\lambda6716/\lambda6731$ measurements \citep{Moustakas2006}, whereas the far-infrared densities were based on [O\,{\sc iii}] 52~$\micron$/88~$\micron$ measurements obtained within the fixed $\sim75\arcsec$ ISO/LWS beam \citep{Brauher2008}. For spatially extended nearby galaxies, these measurements need not sample the same line-emitting regions, complicating direct comparisons.\footnote{Several galaxies in the nearby comparison sample of \citet{Harikane2025}, including NGC~2146, NGC~4038, NGC~1569, and NGC~278, have optical major-axis diameters substantially larger than the $\sim75\arcsec$ ISO/LWS beam (table~1 in \citealt{Brauher2008}).}

Compact, metal-poor dwarf galaxies provide a particularly useful benchmark because they resemble several key properties of high-redshift star-forming systems while allowing optical and far-infrared measurements to encompass most or all of the line-emitting galaxy. Many also show elevated [O\,{\sc iii}] 88~$\micron$/[C\,{\sc ii}] 158~$\micron$ ratios similar to those observed at $z\approx6$--10 \citep[e.g.,][]{Hashimoto2019,Harikane2020,Bakx2024,Algera2024}. Their small angular sizes therefore reduce aperture mismatches and enable more controlled tests of whether different ionized-gas diagnostics can be represented by a single set of physical conditions.

In this study, we test the consistency between optical and far-infrared diagnostics of ionized gas in nearby metal-poor dwarf galaxies selected from the \textit{Herschel} Dwarf Galaxy Survey \citep[HDGS;][]{Cormier2012,Madden2013,Cormier2015,Cormier2019,Ramambason2022}. We combine [O\,{\sc iii}] $\lambda4363$ and $\lambda5007$ measurements with \textit{Herschel}/PACS observations of [O\,{\sc iii}] 88~$\micron$, using optical data that encompass the full galaxy or its dominant line-emitting region. The optical measurements are drawn from our observations with KOOLS-IFU (Kyoto Okayama Optical Low-dispersion Spectrograph with optical-fiber Integral Field Unit) on the 3.8~m Seimei Telescope \citep{Matsubayashi2019,Matsubayashi2025} and from published integral-field spectroscopy. We first use the [O\,{\sc iii}] $\lambda4363/\lambda5007$ and 88~$\micron/\lambda5007$ ratios to test the one-zone description and infer the effective electron density implied by the optical--far-infrared emission. We then compare this electron density with independent optical density diagnostics and explore whether unresolved mixtures of relatively dense and diffuse gas can reproduce the observed [O\,{\sc iii}] emission.

The remainder of this paper is organized as follows. Section~\ref{subsec:obs2} describes the sample, optical spectroscopy, ancillary data, and far-infrared line measurements. Section~\ref{sec:results3} presents the optical--far-infrared [O\,{\sc iii}] diagnostics, the inferred electron temperatures and densities, and representative two-zone models for galaxies showing strong density discrepancies. Section~\ref{sec:discussion4} discusses the diversity of ionized-gas structures in nearby metal-poor galaxies and the implications for high-redshift observations. Section~\ref{sec:conclusions5} summarizes our conclusions.

We adopt a flat $\Lambda$CDM cosmology consistent with \citet{Planck2020}, with $H_0=67.4$~km~s$^{-1}$~Mpc$^{-1}$, $\Omega_{\mathrm{m}}=0.315$, and $\Omega_{\Lambda}=0.685$. We assume a solar luminosity of $L_\odot=3.842\times10^{33}$~erg~s$^{-1}$. Throughout this paper, rest-frame optical emission-line wavelengths are given in air.

\begin{table*}[!htbp]
\tbl{Basic properties and far-infrared line measurements of the primary sample.}{%
\centering
\begin{tabular}{lccccc}
\hline
Quantity & HS 1222+3741 & SBS 0335--052E & POX 186 & Haro 11 & I Zw 18 \\
\hline\hline
RA (J2000) & 12h24m36.7s & 03h37m44.1s & 13h25m48.7s & 00h36m52.5s & 09h34m02.0s \\
Dec (J2000) & +37d24m37s & $-$05d02m40s & $-$11d36m38s & $-$33d33m19s & +55d14m28s \\
Redshift & 0.0409 & 0.0135 & 0.0039 & 0.0206 & 0.0025 \\
$\log(M_\ast/M_\odot)$ & 9.37 & 8.03 & 7.08 & 10.24 & 7.34 \\
$12+\log({\rm O/H})$ & 7.79 & 7.25 & 7.70 & 8.36 & 7.14 \\
$L_{\rm [OIII]88}/L_{\rm [CII]158}$ & $6.0 \pm 1.3$ & $6.6 \pm 1.0$ & $10.7 \pm 3.0$ & $2.63 \pm 0.05$ & $2.68 \pm 0.67$ \\
$F_{\rm [OIII]88}$ & $1.46 \pm 0.15$ & $4.36 \pm 0.26$ & $3.37 \pm 0.33$ & $172 \pm 3$ & $2.84 \pm 0.34$ \\
$F_{\rm [CII]158}$ & $0.24 \pm 0.05$ & $0.66 \pm 0.10$ & $0.31 \pm 0.08$ & $65.5 \pm 0.7$ & $1.06 \pm 0.23$ \\
\hline
\end{tabular}
}
\begin{tabnote}
\begin{minipage}[t]{0.7\textwidth}
\raggedright
{\bf Notes:}
Coordinates, redshifts, stellar masses, metallicities, and far-infrared line fluxes are adopted from \citet{Cormier2015}. The line fluxes are given in units of $10^{-17}$ W m$^{-2}$. The $L_{\rm [OIII]88}/L_{\rm [CII]158}$ ratios are calculated directly from the listed line fluxes.
\end{minipage}
\end{tabnote}
\label{tab:sample}
\end{table*}

\section{Observations and Data}
\label{subsec:obs2}

This section describes the sample selection and the optical and far-infrared spectroscopic data used in this study.

\subsection{Sample Selection}
\label{subsec:sample2.1}

Our parent sample is the HDGS, which comprises 50 nearby low-metallicity dwarf galaxies spanning $12+\log({\rm O/H})=7.14$--8.43. Most HDGS galaxies have \textit{Herschel}/PACS spectroscopy of [O\,{\sc iii}] 88~$\micron$ and [C\,{\sc ii}] 158~$\micron$. The [O\,{\sc iii}] 52~$\micron$ line lies near the short-wavelength limit of PACS and was not measured in the HDGS observations presented by \citet{Cormier2015}; therefore, the [O\,{\sc iii}] 52~$\micron$/88~$\micron$ density diagnostic is unavailable for our sample.

From the parent sample, we first selected 41 galaxies with galaxy-integrated measurements of both [C\,{\sc ii}] 158~$\micron$ and [O\,{\sc iii}] 88~$\micron$. We then required (1) declination larger than $-25^\circ$, ensuring observability with the Seimei Telescope; (2) an angular extent comparable to or smaller than the KOOLS-IFU field of view, $8\farcs4 \times 8\farcs0$; and (3) $L_{\rm [OIII]88}/L_{\rm [CII]158}\gtrsim2$, selecting galaxies with elevated far-infrared line ratios similar to those observed in high-redshift star-forming systems and thus particularly relevant as local benchmarks \citep[e.g.,][]{Hashimoto2018, Hashimoto2019,Hashimoto2019b,Ura2023,Bakx2024,Algera2024}. These criteria yielded five candidates: HS~0017+1055, HS~1222+3741, HS~1319+3224, POX~186, and SBS~0335--052E.

Seimei/KOOLS-IFU observations were successfully obtained for three candidates: HS~1222+3741, SBS~0335--052E, and POX~186. To our knowledge, these observations provide the first optical integral-field spectroscopy of HS~1222+3741. Although POX~186 was observed with KOOLS-IFU, we adopt the higher-quality galaxy-integrated optical emission-line fluxes reported by \citet{Kumari2024}.

To enlarge the sample, we performed an additional search of the HDGS. Because the criteria above were designed specifically for new Seimei/KOOLS-IFU observations, we repeated the selection using only the far-infrared line-ratio criterion, without imposing constraints on declination or angular extent. Among these galaxies, we searched for systems with suitable published or archival optical integral-field spectroscopy and identified Haro~11, Haro~3, and I~Zw~18.

Because the optical data for Haro~3 cover only Region~A, one of the galaxy's principal star-forming regions, whereas the far-infrared measurement is galaxy-integrated \citep{Chen2026}, we include Haro~3 only as a supplementary comparison object. The primary sample therefore comprises five galaxies: HS~1222+3741, SBS~0335--052E, POX~186, Haro~11, and I~Zw~18. The Seimei/KOOLS-IFU observations and ancillary optical spectroscopy are described in sections~\ref{subsec:seimei2.2} and \ref{subsec:ancillary2.3}, respectively. For all five galaxies, we adopt the far-infrared [O\,{\sc iii}] 88~$\micron$ and [C\,{\sc ii}] 158~$\micron$ fluxes reported by \citet{Cormier2015}. Table~\ref{tab:sample} summarizes the basic galaxy properties and the far-infrared emission-line fluxes used in this study.

\begin{table*}[t]
\tbl{Log of the Seimei/KOOLS-IFU observations.}{%
\centering
\begin{tabular}{lcccccc}
\hline
Target & Date & Grism & Spectral range & Resolving power & Exposure time & Standard stars \\
       &      &       & [\AA]          &                 & [s]           &                \\
\hline\hline
HS 1222+3741   & 2021 Dec. 7 & VPH683   & 5800--8000 & $R\sim2000$ & 1800 & BD+75d325 \\
               & 2021 Dec. 8 & VPH495   & 4300--5900 & $R\sim1500$ & 3600 & HR~1544, BD+75d325 \\
SBS 0335--052E & 2021 Dec. 6 & VPH-blue & 4100--8900 & $R\sim600$  & 1500 & HD~15318 \\
POX 186        & 2022 Mar. 7 & VPH683   & 5800--8000 & $R\sim2000$ & 7200 & HR~5501, HR~4963 \\
               & 2022 Mar. 7 & VPH495   & 4300--5900 & $R\sim1500$ & 1200 & HR~5501 \\
\hline
\end{tabular}
}
\label{tab:obslog}
\end{table*}

\begin{table*}[t]
\tbl{Adopted optical emission-line fluxes.}{%
\centering
\begin{tabular}{lccccc}
\hline
Line & HS 1222+3741 & SBS 0335--052E & POX 186 & Haro 11 & I Zw 18 \\
\hline\hline
{}[O\,{\sc ii}] $\lambda3726$ & $\cdots$ & $\cdots$ & $\cdots$ & $139.6 \pm 9.74$ & $\cdots$ \\
{}[O\,{\sc ii}] $\lambda3729$ & $\cdots$ & $\cdots$ & $\cdots$ & $166.7 \pm 12.56$ & $\cdots$ \\
{}[O\,{\sc iii}] $\lambda4363$ & $<0.856\,(3\sigma)$ & $1.49 \pm 0.05$ & $0.65 \pm 0.11$ & $2.46 \pm 0.25$ & $1.00 \pm 0.10$ \\
H$\beta$ & $1.80 \pm 0.06$ & $14.49 \pm 0.02$ & $4.81 \pm 0.05$ & $102.6 \pm 0.6$ & $15.90 \pm 0.11$ \\
{}[O\,{\sc iii}] $\lambda5007$ & $9.9 \pm 1.0$ & $41.76 \pm 0.08$ & $30.8 \pm 1.2$ & $348.1 \pm 2.7$ & $28.62 \pm 0.33$ \\
{}[S\,{\sc iii}] $\lambda6312$ & $\cdots$ & $0.08 \pm 0.007$ & $\cdots$ & $\cdots$ & $\cdots$ \\
H$\alpha$ & $5.04 \pm 0.07$ & $39.83 \pm 0.05$ & $13.76 \pm 0.43$ & $296.3 \pm 1.7$ & $43.60 \pm 0.32$ \\
{}[S\,{\sc ii}] $\lambda6716$ & $\cdots$ & $0.29 \pm 0.03$ & $0.175 \pm 0.008$ & $53.6 \pm 1.0$ & $0.65 \pm 0.03$ \\
{}[S\,{\sc ii}] $\lambda6731$ & $\cdots$ & $0.21 \pm 0.03$ & $0.139 \pm 0.009$ & $56.4 \pm 1.0$ & $0.45 \pm 0.05$ \\
{}[S\,{\sc iii}] $\lambda9069$ & $\cdots$ & $0.45 \pm 0.008$ & $\cdots$ & $\cdots$ & $\cdots$ \\
\hline
\end{tabular}
}
\begin{tabnote}
\begin{minipage}[t]{0.7\textwidth}
\raggedright
{\bf Notes:}
Optical emission-line fluxes are given in units of $10^{-14}$ erg s$^{-1}$ cm$^{-2}$. For HS~1222+3741, the fluxes were measured from our KOOLS-IFU observations. For SBS~0335--052E, the H$\beta$, [O\,{\sc iii}] $\lambda5007$, [S\,{\sc iii}] $\lambda6312$, H$\alpha$, [S\,{\sc ii}] $\lambda\lambda6716,6731$, and [S\,{\sc iii}] $\lambda9069$ fluxes were measured from the galaxy-integrated MUSE spectrum. The [O\,{\sc iii}] $\lambda4363$ flux was measured from our KOOLS-IFU data and multiplied by an aperture-correction factor of 1.3 (section~\ref{subsubsec:ancillary-SBS2.3.1}). For POX~186 and I~Zw~18, the fluxes were adopted from \citet{Kumari2024} and \citet{Kehrig2016}, respectively. For Haro~11, the [O\,{\sc ii}] $\lambda\lambda3726,3729$ and [O\,{\sc iii}] $\lambda4363$ fluxes were adopted from \citet{James2013} and multiplied by an aperture-correction factor of 2.0, while the H$\beta$, [O\,{\sc iii}] $\lambda5007$, H$\alpha$, and [S\,{\sc ii}] $\lambda\lambda6716,6731$ fluxes were measured from the galaxy-integrated MUSE spectrum (section~\ref{subsubsec:ancillary-Haro11.2.3.3}).
\end{minipage}
\end{tabnote}
\label{tab:lineflux}
\end{table*}

\begin{figure*}[!htbp]
    \centering
    \includegraphics[width=0.4\linewidth]{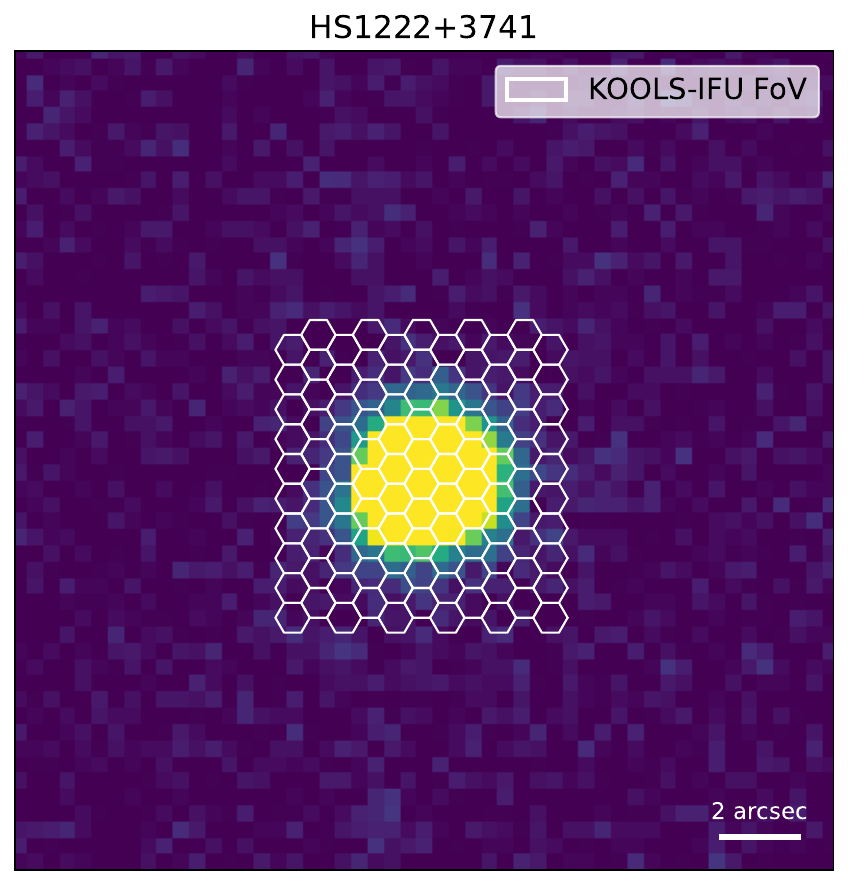}
    \hspace{0.8cm}
    \includegraphics[width=0.4\linewidth]{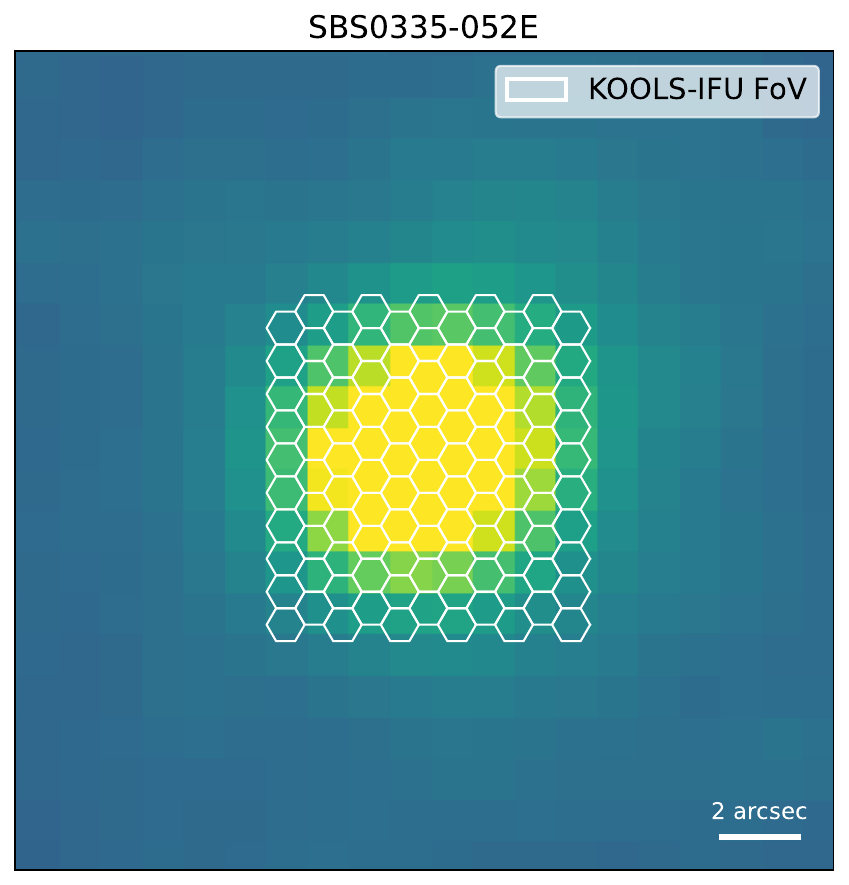}
\caption{
\raggedright
KOOLS-IFU fields of view overlaid on optical images of HS~1222+3741 (left) and SBS~0335--052E (right). The 110 hexagons represent the science fibers sampling the KOOLS-IFU field of view. The background images are an SDSS DR7 $g$-band image of HS~1222+3741 and a DSS blue-band image of SBS~0335--052E, obtained via NED. The scale bars at the bottom right of each panel indicate $2\arcsec$. 
%
}
    \label{fig:seimei_FoV}
\end{figure*}

\begin{figure*}[!htbp]
    \centering
    \includegraphics[width=12cm]{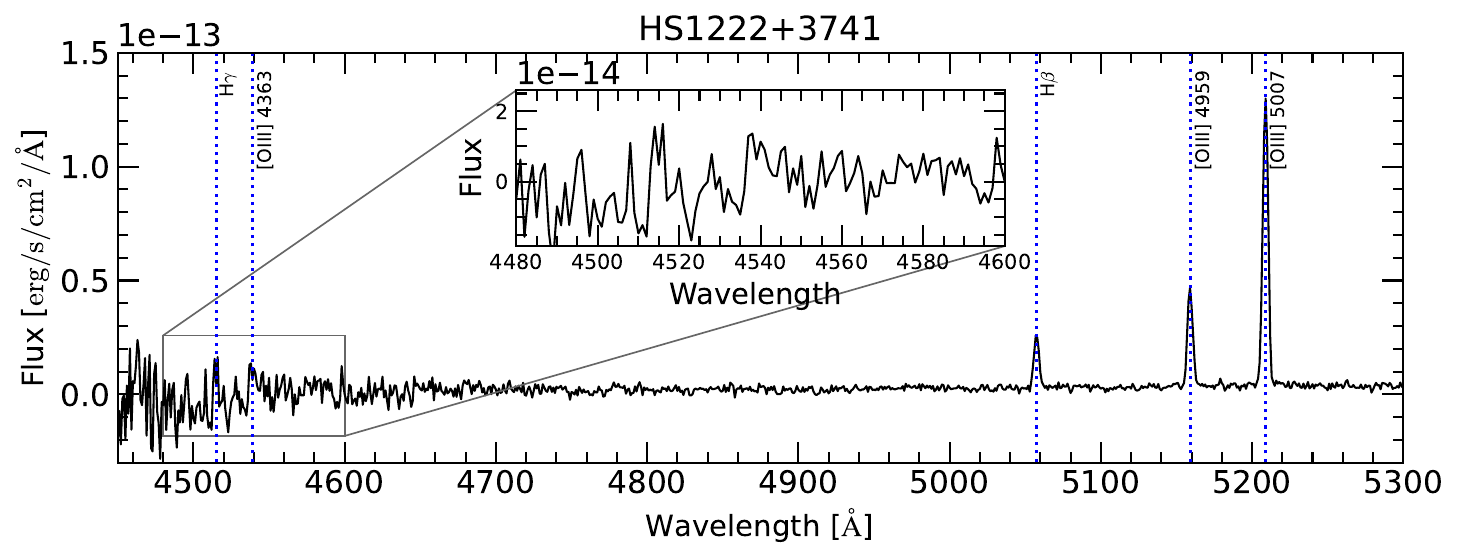}
    \hspace{0.2cm}
    \includegraphics[width=4.5cm]{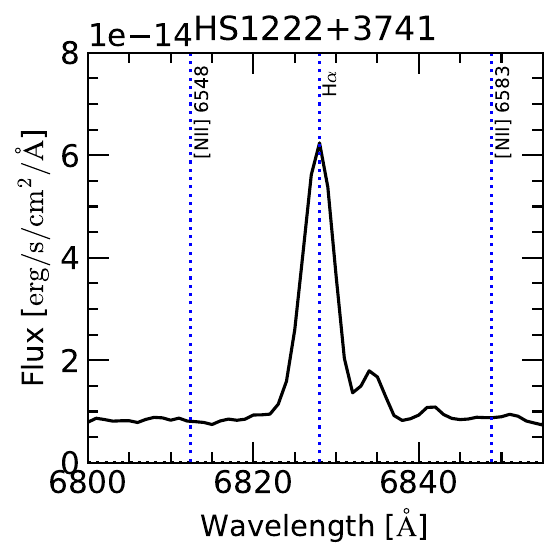}
    \vspace{0.4cm}
    \includegraphics[width=15cm]{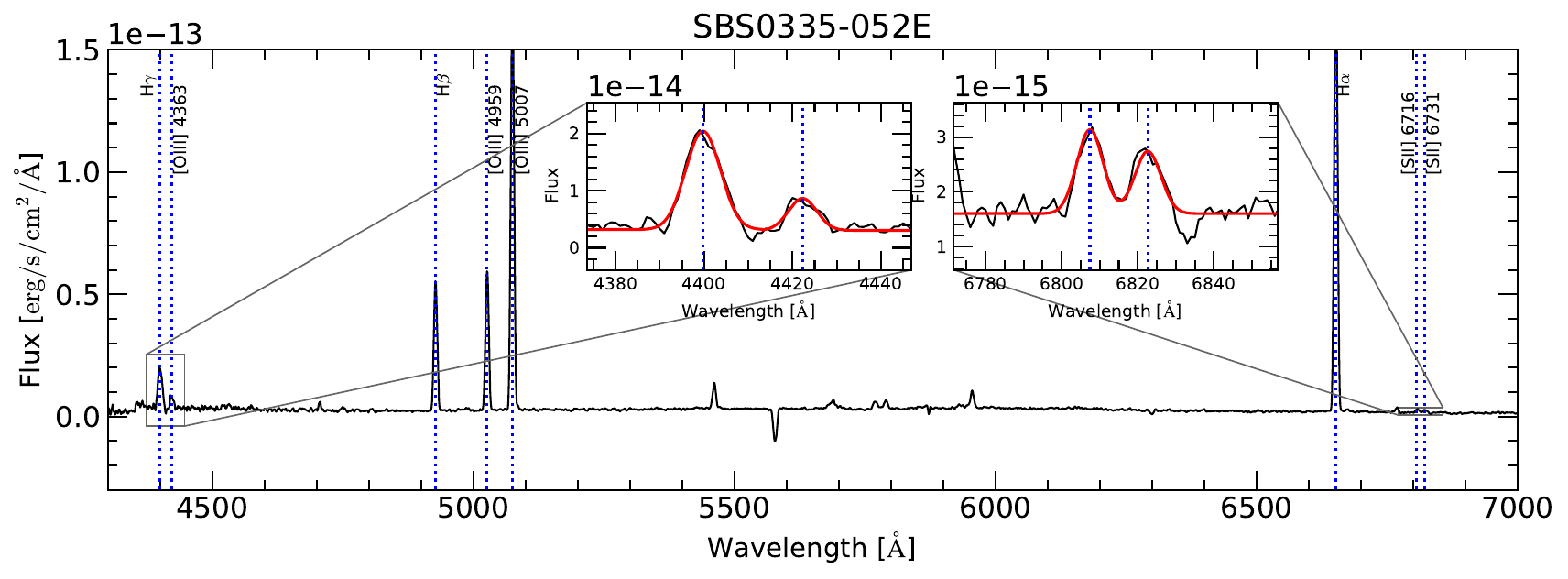}
\caption{
\raggedright
Seimei/KOOLS-IFU spectra of HS~1222+3741 and SBS~0335--052E. Top left: HS~1222+3741 observed with the VPH495 grism. Top right: HS~1222+3741 observed with the VPH683 grism. Bottom: SBS~0335--052E observed with the VPH-blue grism. The horizontal axes show the observed-frame wavelength. The vertical blue dashed lines indicate the expected wavelengths of the emission lines. The zoomed-in inset in the top-left panel shows the spectral regions around H$\gamma$ and [O\,{\sc iii}] $\lambda4363$; neither line is detected in HS~1222+3741. The zoomed-in insets in the bottom panel show the spectral regions around H$\gamma$ and [O\,{\sc iii}] $\lambda4363$, and around the [S\,{\sc ii}] $\lambda\lambda6716,6731$ doublet, respectively; these emission lines are detected in SBS~0335--052E. The red curves show Gaussian fits to the emission lines.
%
}
    \label{fig:Obsspec}
\end{figure*}

\subsection{Seimei/KOOLS-IFU Observations and Data Reduction}
\label{subsec:seimei2.2}

We conducted optical integral-field spectroscopic observations of HS~1222+3741 and SBS~0335--052E during the 2021B semester, and of POX~186 during the 2022A semester, with KOOLS-IFU \citep{Matsubayashi2019,Matsubayashi2025} on the 3.8~m Seimei Telescope at Okayama Observatory (PI: T. Hashimoto; proposal IDs: 21B-N-CN07 and 22A-N-CN07). KOOLS-IFU is an optical fiber-fed integral-field spectrograph with 117 fibers, of which 110 sample the science field and seven are assigned to blank-sky regions, covering a field of view of $8\farcs4 \times 8\farcs0$. The seeing ranged from $1\farcs2$ to $1\farcs4$. The observations were obtained under clear conditions. The observing log is summarized in table~\ref{tab:obslog}. Figure~\ref{fig:seimei_FoV} shows the KOOLS-IFU fields of view for HS~1222+3741 and SBS~0335--052E, overlaid on an SDSS DR7 $g$-band image and a DSS blue-band image, respectively, obtained via the NASA/IPAC Extragalactic Database (NED).\footnote{\url{https://ned.ipac.caltech.edu/}}

HS~1222+3741 was observed with the VPH495 and VPH683 grisms. The VPH495 observations were obtained on 2021 December 8, covering 4300--5900~\AA\ at a spectral resolving power of $R\sim1500$. This wavelength range includes [O\,{\sc iii}] $\lambda4363$, H$\beta$, and [O\,{\sc iii}] $\lambda5007$. The total on-source exposure time was 3600~s, consisting of five 600~s exposures and two 300~s exposures.
The VPH683 observations were obtained on 2021 December 7, covering 5800--8000~\AA\ at $R\sim2000$. This wavelength range includes H$\alpha$ and the [S\,{\sc ii}] $\lambda\lambda6716,6731$ doublet. The total on-source exposure time was 1800~s, consisting of three 600~s exposures.

SBS~0335--052E was observed on 2021 December 6 with the VPH-blue grism, covering 4100--8900~\AA\ at a spectral resolving power of $R\sim600$. This wavelength range includes [O\,{\sc iii}] $\lambda4363$, H$\beta$, [O\,{\sc iii}] $\lambda5007$, H$\alpha$, and the [S\,{\sc ii}] $\lambda\lambda6716,6731$ doublet. The total on-source exposure time was 1500~s, consisting of two 600~s exposures and one 300~s exposure.

POX~186 was observed on 2022 March 7 with the VPH495 and VPH683 grisms. The VPH495 observations covered 4300--5900~\AA\ at a spectral resolving power of $R\sim1500$, including [O\,{\sc iii}] $\lambda4363$, H$\beta$, and [O\,{\sc iii}] $\lambda5007$. The total on-source exposure time was 1200~s, consisting of two 600~s exposures. The VPH683 observations covered 5800--8000~\AA\ at $R\sim2000$, including H$\alpha$ and the [S\,{\sc ii}] $\lambda\lambda6716,6731$ doublet. The total on-source exposure time was 7200~s, consisting of twelve 600~s exposures.

Data reduction was carried out with tasks in the Image Reduction and Analysis Facility (\texttt{IRAF}; \citealt{Tody1986,Tody1993}) accessed through the Python interface \texttt{PyRAF} \citep{PyRAF2012}. The procedure included overscan and bias subtraction, bad-column correction, flat-fielding, wavelength calibration, spectral extraction, sky subtraction, and flux calibration. Hg, Ne, and Xe lamp exposures were used for wavelength calibration, and dedicated sky frames were used for sky subtraction. Flux calibration was based on the spectrophotometric standard stars listed in table~\ref{tab:obslog}, observed on the same nights and at airmasses similar to those of the corresponding science targets.

The standard KOOLS-IFU reduction does not explicitly correct for atmospheric extinction. However, because the spectrophotometric standard stars were observed at airmasses similar to those of the science targets, the residual differential atmospheric-extinction effect on the relative line fluxes is expected to be small. The wavelength-dependent sensitivity was corrected using the sensitivity function derived from the standard-star observations. Narrow telluric absorption features are not explicitly removed by this procedure because the sensitivity function is fitted with a low-order polynomial, typically of order 6--12.

For each target, we inspected the one-dimensional spectrum extracted from each fiber and identified fibers showing clear nebular emission at the expected wavelengths. We selected the contiguous group of such fibers associated with the target and excluded fibers dominated by noise or without significant line emission. We then examined the spatial distributions of [O\,{\sc iii}] $\lambda5007$ and H$\alpha$ to confirm that the selected fibers encompass the dominant line-emitting regions. Spatially integrated spectra were constructed by summing the selected fiber spectra. The resulting spectra are shown in figure~\ref{fig:Obsspec}.

Emission-line fluxes were measured by fitting each line with a single Gaussian profile plus a local continuum using \texttt{curve\_fit} in \texttt{scipy}. The line flux was obtained by integrating the fitted Gaussian component. Flux uncertainties were estimated from 1000 Monte Carlo realizations, in which Gaussian noise with a standard deviation estimated from the local continuum was added to the observed spectrum. Each realization was then fitted in the same manner as the original spectrum. We adopted the median of the resulting flux distribution as the measured flux and the 16th--84th percentile range as the $1\sigma$ uncertainty. Lines with $S/N\geq3$ were regarded as detections.

The measured optical emission-line fluxes were corrected for Galactic foreground extinction using the $E(B-V)_{\rm MW}$ values obtained from the IRSA Galactic Dust Reddening and Extinction Service.\footnote{\url{https://irsa.ipac.caltech.edu/applications/DUST/}} These values are based on the dust maps of \citet{Schlegel1998} recalibrated by \citet{Schlafly2011}. We adopted a Milky Way extinction curve with $R_V=3.1$ \citep{Cardelli1989}. After correcting for Galactic foreground extinction, we used the H$\alpha$/H$\beta$ ratios to assess the internal dust attenuation in each galaxy, as described below.

For HS~1222+3741, we adopted $E(B-V)_{\rm MW}=0.01$, corresponding to $A_{V,\rm MW}=0.03$~mag. After correction for Galactic foreground extinction, the measured H$\alpha$/H$\beta$ ratio is $2.78\pm0.02$, consistent with the dust-free Case~B recombination ratio of 2.77 at $T_{\rm e}=18,000~{\rm K}$ and $n_{\rm e}=100~{\rm cm}^{-3}$ \citep{Osterbrock2006}. We therefore applied no additional correction for internal dust attenuation.

For SBS~0335--052E, we adopted $E(B-V)_{\rm MW}=0.04$, corresponding to $A_{V,\rm MW}=0.1$~mag. We assessed the internal dust attenuation using the higher-quality archival MUSE measurements of H$\alpha$ and H$\beta$ \citep{Herenz2017}. After correction for Galactic foreground extinction, the resulting H$\alpha$/H$\beta$ ratio is $2.86\pm0.01$. Adopting the dust-free Case~B recombination ratio of 2.75 appropriate for $T_{\rm e}=20,000~{\rm K}$ and $n_{\rm e}=100~{\rm cm}^{-3}$ \citep{Osterbrock2006} (see section~\ref{subsec:results3.2}), we obtain $E(B-V)_{\rm int}=0.04\pm0.01$ assuming an SMC extinction curve \citep{Gordon2003}. We applied this internal attenuation correction to the SBS~0335--052E optical emission-line fluxes used in the subsequent analysis.

The corrected Seimei/KOOLS-IFU fluxes are adopted for HS~1222+3741 and combined with archival optical data for SBS~0335--052E, as described in the following subsection. For POX~186, the KOOLS-IFU measurements serve only as an independent check of the absolute flux calibration; the higher-quality galaxy-integrated fluxes from \citet{Kumari2024} are used for the scientific analysis.

\subsection{Ancillary Optical Spectroscopic Data}
\label{subsec:ancillary2.3}

In addition to the Seimei/KOOLS-IFU observations described above, we used optical integral-field spectroscopic data and emission-line measurements from the literature. For SBS~0335--052E and Haro~11, we analyzed reduced VLT/MUSE data cubes and remeasured the relevant emission-line fluxes using apertures chosen to facilitate comparison with the far-infrared measurements. We also used VLT/FLAMES measurements for Haro~11 and published integral-field spectroscopic measurements for I~Zw~18 and POX~186. Galaxy-integrated or nearly galaxy-integrated measurements were adopted whenever possible to minimize aperture mismatches with the \textit{Herschel}/PACS observations.

Although integral-field spectroscopy provides spatially resolved information, we use spatially integrated line ratios because the \textit{Herschel}/PACS [O\,{\sc iii}] 88~$\micron$ measurements are galaxy-integrated and [O\,{\sc iii}] $\lambda4363$ is generally too faint for uniform spatially resolved analysis across the sample. The IFU data are therefore used primarily to encompass the dominant line-emitting regions and construct optical measurements comparable to the far-infrared fluxes.

\subsubsection{SBS~0335--052E}
\label{subsubsec:ancillary-SBS2.3.1}

For SBS~0335--052E, we measured galaxy-integrated optical emission-line fluxes from the archival VLT/MUSE observations presented by \citet{Herenz2017}. The MUSE observations cover a $1\arcmin \times 1\arcmin$ field of view over 4600--9370~\AA\ and encompass the entire galaxy (see figure~1 of \citealt{Herenz2017}). We measured galaxy-integrated fluxes of H$\beta$, [O\,{\sc iii}] $\lambda5007$, [S\,{\sc iii}] $\lambda6312$, H$\alpha$, [S\,{\sc ii}] $\lambda\lambda6716,6731$, and [S\,{\sc iii}] $\lambda9069$. The [O\,{\sc iii}] $\lambda4363$ line lies outside the MUSE wavelength coverage and was measured from our Seimei/KOOLS-IFU spectrum.

Because the [O\,{\sc iii}] $\lambda4363$ and $\lambda5007$ fluxes are obtained with different instruments, we checked their relative flux calibration by measuring the [O\,{\sc iii}] $\lambda5007$ flux from the MUSE cube within an aperture matched to the KOOLS-IFU field of view. The aperture-matched MUSE and KOOLS-IFU [O\,{\sc iii}] $\lambda5007$ fluxes agree within the uncertainties, supporting the consistency of the relative flux calibration between the two data sets.

Because the KOOLS-IFU field of view does not encompass the entire galaxy, we applied an aperture-correction factor of 1.3 to the measured [O\,{\sc iii}] $\lambda4363$ flux. This factor was derived from the ratio of the galaxy-integrated MUSE [O\,{\sc iii}] $\lambda5007$ flux to that measured within an aperture matched to the KOOLS-IFU field of view. This correction assumes that the [O\,{\sc iii}] $\lambda4363/\lambda5007$ ratio within the KOOLS-IFU field is representative of the galaxy as a whole.

As an independent check of the relative calibration of [O\,{\sc iii}] $\lambda4363$ and $\lambda5007$, we compared our $\lambda4363/\lambda5007$ ratio with the galaxy-integrated VLT/GIRAFFE measurement of \citet{Izotov2006}. Their observed ratio of 0.0319 agrees well with our value of 0.0338, differing by only $\simeq6\%$.

\subsubsection{POX~186}
\label{subsubsec:ancillary-POX2.3.2}

For POX~186, we adopted the optical emission-line fluxes reported by \citet{Kumari2024}, obtained with the Gemini Multi-Object Spectrograph Integral Field Unit (GMOS-IFU) on the Gemini North Telescope. The observations were conducted in one-slit mode, covering $3\farcs5 \times 5\farcs0$ and encompassing the entire galaxy (see figure~2 of \citealt{Kumari2024}). Of the two spatially integrated spectra presented by \citet{Kumari2024}, we used the Gemini-FOV integrated spectrum constructed by summing the full GMOS-IFU field of view.\footnote{As an independent check of the absolute flux calibration, the emission-line fluxes measured from our KOOLS-IFU observations are consistent with the galaxy-integrated GMOS-IFU measurements.}

The published fluxes were corrected for Galactic foreground extinction using $E(B-V)_{\rm MW}=0.04$ and for internal dust attenuation using $E(B-V)_{\rm int}=0.08$, adopting the SMC extinction curve of \citet{Gordon2003}. This internal attenuation is broadly consistent with the independent estimate of \citet{Rogers2023}, who obtained $E(B-V)=0.129\pm0.004$ from UV stellar-population modeling.

\subsubsection{Haro~11}
\label{subsubsec:ancillary-Haro11.2.3.3}

For Haro~11, we combined published VLT/FLAMES measurements with measurements from reduced VLT/MUSE data. We adopted the [O\,{\sc ii}] $\lambda\lambda3726,3729$ and [O\,{\sc iii}] $\lambda4363$ fluxes reported by \citet{James2013}, because these lines lie outside the wavelength coverage of the MUSE data used here.

We measured galaxy-integrated H$\beta$, [O\,{\sc iii}] $\lambda5007$, H$\alpha$, and [S\,{\sc ii}] $\lambda\lambda6716,6731$ fluxes from the VLT/MUSE cube analyzed by \citet{Menacho2021}. The MUSE observations encompass the full optical line-emitting extent of Haro~11 (see figure~1 of \citealt{Menacho2021}). We constructed a galaxy-integrated spectrum and measured the line fluxes using the same fitting procedure as for the Seimei/KOOLS-IFU spectra.

Because the [O\,{\sc ii}] $\lambda\lambda3726,3729$ and [O\,{\sc iii}] $\lambda4363$ measurements are obtained with VLT/FLAMES whereas [O\,{\sc iii}] $\lambda5007$ is measured from VLT/MUSE, we account explicitly for the different spatial coverages of the two data sets. The VLT/FLAMES observations covered $11\farcs5 \times 7\farcs3$ and did not encompass the full galaxy (see figure~1 of \citealt{James2013}). We therefore applied an aperture-correction factor of 2.0 to the published [O\,{\sc ii}] $\lambda\lambda3726,3729$ and [O\,{\sc iii}] $\lambda4363$ fluxes. This factor was derived from the ratio of the galaxy-integrated MUSE [O\,{\sc iii}] $\lambda5007$ flux to that measured within an aperture matched to the VLT/FLAMES field of view. This correction assumes that the spatial distributions of [O\,{\sc ii}] and [O\,{\sc iii}] $\lambda4363$ relative to [O\,{\sc iii}] $\lambda5007$ within the FLAMES field are representative of the galaxy as a whole.

The optical fluxes were corrected for Galactic foreground extinction using $E(B-V)_{\rm MW}=0.01$. From the galaxy-integrated MUSE Balmer decrement, H$\alpha$/H$\beta=3.61\pm0.01$, we derived an internal attenuation of $A_V=0.6$~mag, adopting an intrinsic Case~B ratio of 2.86 for $T_{\rm e}=10,000$~K and $n_{\rm e}=100~{\rm cm}^{-3}$ and the SMC extinction curve of \citet{Gordon2003}. This representative temperature is consistent with the moderately metal-poor nature of Haro~11, with $12+\log({\rm O/H})\simeq8.4$ (see table~\ref{tab:sample}).

\subsubsection{I~Zw~18}
\label{subsubsec:ancillary-IZw2.3.4}

For I~Zw~18, we adopted the optical emission-line fluxes reported by \citet{Kehrig2016}, obtained with the Potsdam Multi-Aperture Spectrophotometer (PMAS) on the 3.5~m telescope at Calar Alto Observatory. The PMAS data cover $16^{\prime\prime} \times 16^{\prime\prime}$, encompassing the entire main body of I~Zw~18 and part of the surrounding extended ionized gas (see figure~1 of \citealt{Kehrig2016}). We used the fluxes from their integrated spectrum, which includes nearly all nebular emission within this area. The published fluxes had already been corrected for Galactic foreground extinction using $A_V=0.09$~mag. The Balmer decrement is consistent with the dust-free Case~B recombination value, and no additional correction for internal dust attenuation was applied.

The optical emission-line fluxes used in the present analysis are summarized in table~\ref{tab:lineflux}.

\begin{figure*}[!htbp]
\centering
\includegraphics[width=0.75\linewidth]{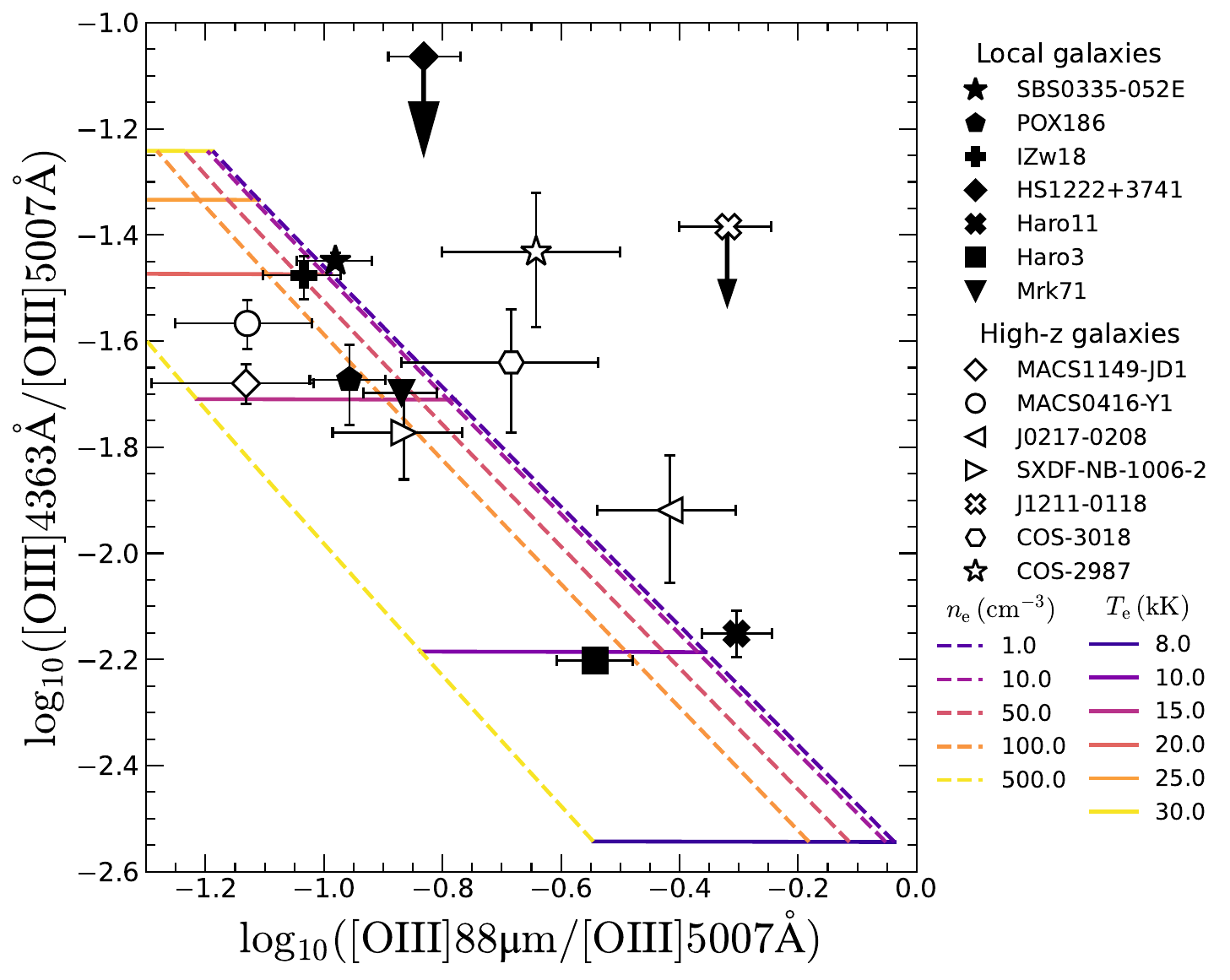}
\caption{
\raggedright
[O\,{\sc iii}] diagnostic diagram based on the [O\,{\sc iii}] 88~$\micron$/[O\,{\sc iii}] $\lambda5007$ and [O\,{\sc iii}] $\lambda4363$/[O\,{\sc iii}] $\lambda5007$ ratios. The horizontal and vertical axes show $\log_{10}(\mbox{[O\,{\sc iii}]}\ 88~\micron/\mbox{[O\,{\sc iii}]}\ \lambda5007)$ and $\log_{10}(\mbox{[O\,{\sc iii}]}\ \lambda4363/\mbox{[O\,{\sc iii}]}\ \lambda5007)$, respectively. Solid and dashed curves show one-zone model predictions calculated with \texttt{PyNeb} version 1.1.32 for different electron temperatures, $T_{\rm e}$, and electron densities, $n_{\rm e}$, respectively. Filled symbols indicate nearby galaxies, including the five galaxies in our primary sample, Mrk~71 \citep{Chen2023}, and Haro~3 Region~A \citep{Chen2026}. Open symbols indicate high-redshift galaxies from the literature \citep{Stiavelli2023,Harshan2024,Fujimoto2024,Usui2025,Harikane2025,Takechi2026}. The downward arrow for HS~1222+3741 indicates the 3$\sigma$ upper limit on [O\,{\sc iii}] $\lambda4363$; this galaxy is not used to assess consistency with the one-zone model. The horizontal error bars include an additional 10\% systematic uncertainty to account for possible relative flux-calibration differences between the optical and \textit{Herschel}/PACS measurements. For display purposes, the plotted position of I~Zw~18 is shifted by $-0.03$~dex along the horizontal axis and by $-0.02$~dex along the vertical axis relative to its measured values.
%
}
\label{fig:OIIIdiagram}
\end{figure*}

\begin{figure*}[!htbp]
\centering
\includegraphics[width=0.98\linewidth]{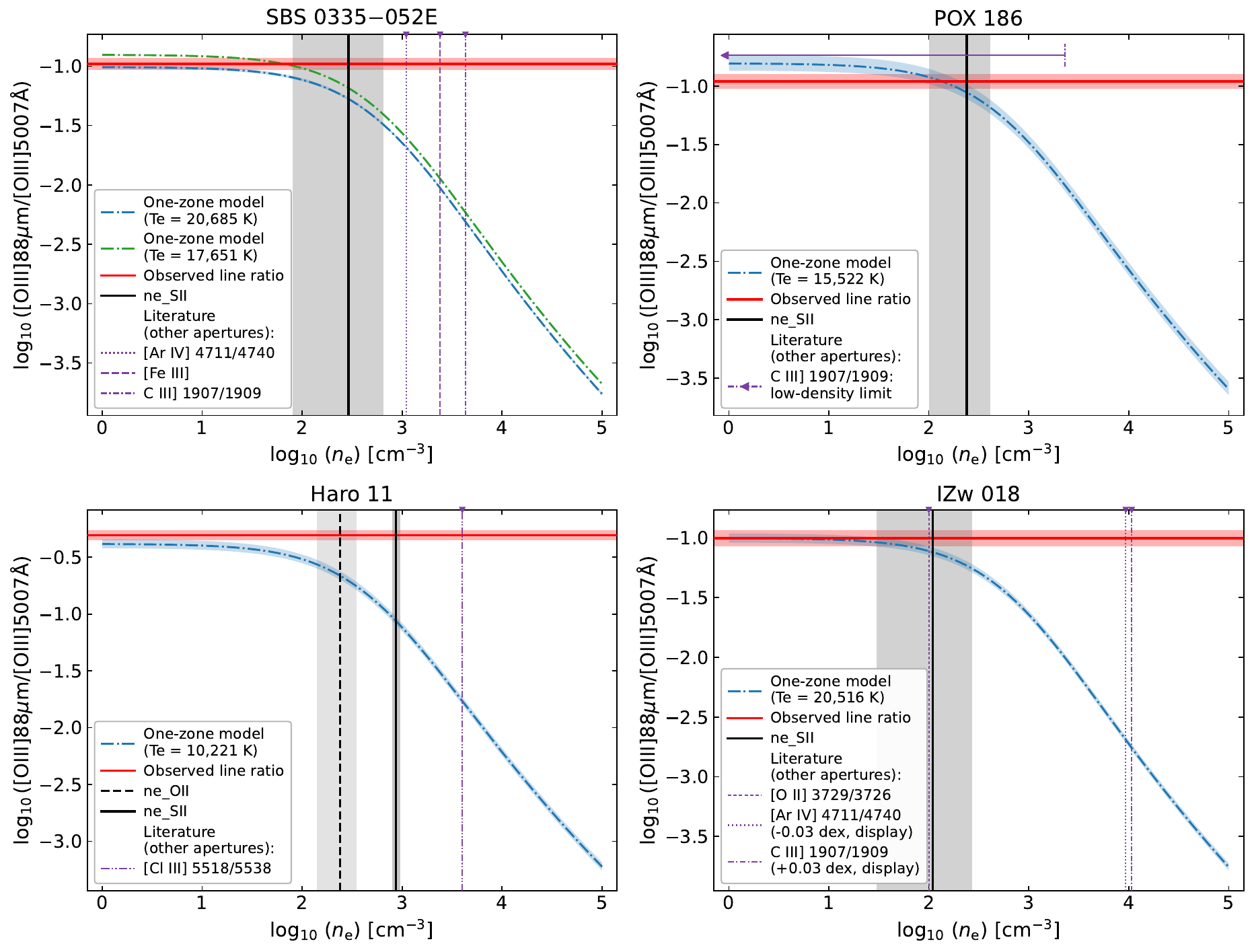}
\caption{
\raggedright
Comparison of electron densities inferred from different optical and ultraviolet diagnostics with the effective electron densities required to reproduce the observed [O\,{\sc iii}] 88~$\micron$/[O\,{\sc iii}] $\lambda5007$ ratios under the one-zone assumption. Top left: SBS~0335--052E. Top right: POX~186. Bottom left: Haro~11. Bottom right: I~Zw~18. For Haro~11, constraints from both the [O\,{\sc ii}] and [S\,{\sc ii}] doublets are shown. The blue curves show the theoretical [O\,{\sc iii}] 88~$\micron/\lambda5007$ ratios as functions of $n_{\rm e}$ at the adopted $T_{\rm e}({\rm O^{++}})$, with the blue shaded regions indicating the uncertainties due to $T_{\rm e}$. For SBS~0335--052E, the green curve additionally shows the prediction adopting $T_{\rm e}({\rm S^{++}})$ derived from [S\,{\sc iii}] $\lambda6312/\lambda9069$. The red horizontal lines and shaded regions show the observed ratios and their $1\sigma$ uncertainties, including the adopted 10\% systematic uncertainty in the optical--far-infrared ratio. The vertical black lines and gray shaded regions indicate the densities and their $1\sigma$ uncertainties inferred from the low-ionization optical doublets; solid and dashed lines denote the [S\,{\sc ii}]- and [O\,{\sc ii}]-based estimates, respectively. Additional literature-based optical and ultraviolet density diagnostics are shown as vertical reference lines and are adopted from \citet{Mingozzi2022} for SBS~0335--052E, Haro~11, and I~Zw~18, and from \citet{Rogers2023} for POX~186. For POX~186, the leftward arrow indicates that the [C\,{\sc iii}] $\lambda1907$/C\,{\sc iii}] $\lambda1909$ ratio is consistent with the low-density limit, corresponding to $n_{\rm e}=1^{+2300}_{-1}~{\rm cm}^{-3}$; the vertical cap marks the $1\sigma$ upper bound.
%
}
\label{fig:Tene}
\end{figure*}

\section{Results}
\label{sec:results3}

We test whether the observed optical and far-infrared [O\,{\sc iii}] emission can be represented by a one-zone ionized-gas model. In section~\ref{subsec:results3.1}, we compare the observed [O\,{\sc iii}] $\lambda4363/\lambda5007$ and 88~$\micron/\lambda5007$ ratios with model predictions assuming a single $T_{\rm e}$ and $n_{\rm e}$. In section~\ref{subsec:results3.2}, we derive $T_{\rm e}({\rm O^{++}})$ from [O\,{\sc iii}] $\lambda4363/\lambda5007$, infer the effective electron density implied by [O\,{\sc iii}] 88~$\micron/\lambda5007$, and compare it with independent optical and ultraviolet electron-density diagnostics. In section~\ref{subsec:results3.3}, we examine whether representative two-zone models combining relatively dense and diffuse low-density ionized gas can reproduce the three observed [O\,{\sc iii}] lines in SBS~0335--052E and Haro~11.

These analyses probe complementary aspects of ionized-gas structure. The [O\,{\sc iii}] $\lambda4363/\lambda5007$ and 88~$\micron/\lambda5007$ diagram tests whether the O$^{++}$-emitting gas can be represented by a single $T_{\rm e}$ and $n_{\rm e}$ and whether the observed ratios approach the low-density limit. The comparison with independent electron-density diagnostics then tests whether the effective electron density implied by the optical--far-infrared [O\,{\sc iii}] emission is consistent with other ionized-gas tracers. The two-zone modeling explores whether a mixture of relatively dense and diffuse low-density gas can reproduce the observed [O\,{\sc iii}] line ratios. Differences among electron-density diagnostics may arise from density stratification between different ionization zones or because individual diagnostics preferentially weight different parts of a broad underlying density distribution \citep{MendezDelgado2026}.

\subsection{[O\,{\sc iii}] Diagnostic Diagram}
\label{subsec:results3.1}

Figure~\ref{fig:OIIIdiagram} shows the [O\,{\sc iii}] diagnostic diagram based on the [O\,{\sc iii}] 88~$\micron$/[O\,{\sc iii}] $\lambda5007$ and [O\,{\sc iii}] $\lambda4363$/[O\,{\sc iii}] $\lambda5007$ ratios. Model grids were calculated with \texttt{PyNeb} version 1.1.32 \citep{Luridiana2015} for $T_{\rm e}=8{,}000$--$30{,}000$~K and $n_{\rm e}=1$, 10, 50, 100, and $500~{\rm cm}^{-3}$, assuming that all three lines arise from a homogeneous component with a single $T_{\rm e}$ and $n_{\rm e}$.\footnote{We adopted the default O$^{++}$ atomic data in \texttt{PyNeb} version 1.1.32: transition probabilities from \citet{FroeseFischer2004}, supplemented by those for the $^1S_0$ level from \citet{StoreyZeippen2000}, energy levels from NIST, and effective collision strengths from \citet{Storey2014}.} These calculations are atomic-level population calculations rather than photoionization models: $T_{\rm e}$ and $n_{\rm e}$ are specified directly, and no ionizing spectrum, ionization parameter, escape fraction, gas abundance pattern, dust content, or depletion factor is assumed. Because all three transitions arise from O$^{++}$, the O$^{++}$ abundance cancels in the line ratios. The optical ratio is primarily sensitive to $T_{\rm e}$, whereas the far-infrared-to-optical ratio depends on both $T_{\rm e}$ and $n_{\rm e}$.

For the [O\,{\sc iii}] 88~$\micron/\lambda5007$ ratio, we include a 10\% systematic uncertainty to account for possible differences in the absolute flux calibration between the optical and far-infrared measurements, in addition to the statistical uncertainties. No additional systematic uncertainty is imposed on the [O\,{\sc iii}] $\lambda4363/\lambda5007$ ratio.

The four filled symbols corresponding to galaxies in our primary sample with robust [O\,{\sc iii}] $\lambda4363$ measurements are used for the quantitative one-zone analysis. SBS~0335--052E and Haro~11 lie at or slightly beyond the low-electron-density boundary of the model grid, whereas POX~186 and I~Zw~18 lie within it. HS~1222+3741 is shown separately as a 3$\sigma$ upper limit because [O\,{\sc iii}] $\lambda4363$ is not detected and is therefore not used to assess consistency with the one-zone model.

Because the offsets of SBS~0335--052E and Haro~11 from the model boundary are small, their positions relative to the grid alone do not provide decisive evidence against a one-zone interpretation. Their nominal line ratios lie beyond the $n_{\rm e}=1~{\rm cm}^{-3}$ model boundary and therefore favor effective electron densities of $n_{\rm e}<1~{\rm cm}^{-3}$. As a conservative estimate of the effect of the observational uncertainties, we also consider the lower $1\sigma$ limits of both line ratios and determine the electron densities allowed by their intersections with the model grid. This gives upper limits of approximately $n_{\rm e}\lesssim40~{\rm cm}^{-3}$ for SBS~0335--052E and $n_{\rm e}\lesssim10~{\rm cm}^{-3}$ for Haro~11. 
As shown in section~\ref{subsec:results3.2}, these conservative upper limits remain substantially below the electron densities inferred from independent density diagnostics at optical and ultraviolet wavelengths.

The filled downward triangle and square show two nearby comparison systems from the literature: Mrk~71 at a distance of 3.4~Mpc \citep{Chen2023} and Region~A in Haro~3 at $z=0.0032$ \citep{Chen2026}, respectively. In both cases, integral-field spectroscopy was used to minimize aperture mismatches between the optical and far-infrared measurements.

The tendency of SBS~0335--052E and Haro~11 toward the low-electron-density regime may indicate a substantial contribution from diffuse, low-density O$^{++}$ gas to the [O\,{\sc iii}] 88~$\micron$ emission. Similar behavior is also observed in some high-redshift galaxies \citep{Usui2025,Harikane2025}, indicating that such diagnostic behavior is not restricted to the early Universe.

\subsection{Electron Densities Inferred from Different Diagnostics}
\label{subsec:results3.2}

We next compare the effective electron densities inferred from the optical--far-infrared [O\,{\sc iii}] emission with those obtained from independent density-sensitive line ratios. For each galaxy with a detected [O\,{\sc iii}] $\lambda4363$ line, we first determine $T_{\rm e}({\rm O^{++}})$ from the [O\,{\sc iii}] $\lambda4363/\lambda5007$ ratio and then infer the effective electron density from the [O\,{\sc iii}] 88~$\micron/\lambda5007$ ratio at the resulting temperature. Figure~\ref{fig:Tene} compares the observed [O\,{\sc iii}] 88~$\micron/\lambda5007$ ratios with the predicted ratios as a function of $n_{\rm e}$ and shows independent electron-density constraints from optical and ultraviolet diagnostics. Table~\ref{tab:tene_comparison} summarizes the electron densities inferred from the different diagnostics. HS~1222+3741 is not included in this quantitative comparison because [O\,{\sc iii}] $\lambda4363$ is not detected.

The [S\,{\sc ii}] and [O\,{\sc ii}] measurements used for our primary comparison are based on galaxy-integrated or nearly galaxy-integrated optical measurements. By contrast, the additional optical and ultraviolet density diagnostics adopted from the literature generally probe different apertures. We therefore use the latter as complementary indicators of the range of electron densities present in each galaxy rather than as direct aperture-matched comparisons with the optical--far-infrared [O\,{\sc iii}] measurements.

For SBS~0335--052E, the [O\,{\sc iii}] $\lambda4363/\lambda5007$ ratio gives $T_{\rm e}({\rm O^{++}})\simeq2.0\times10^{4}$~K. At this temperature, the nominal [O\,{\sc iii}] 88~$\micron/\lambda5007$ ratio lies beyond the $n_{\rm e}=1~{\rm cm}^{-3}$ model boundary and therefore favors an effective electron density of $n_{\rm e}<1~{\rm cm}^{-3}$. As described in section~\ref{subsec:results3.1}, adopting the lower $1\sigma$ limits of both [O\,{\sc iii}] ratios as a conservative treatment of the observational uncertainties allows densities up to $n_{\rm e}\lesssim40~{\rm cm}^{-3}$. This remains below the value of $n_{\rm e}=290^{+360}_{-210}~{\rm cm}^{-3}$ inferred from the [S\,{\sc ii}] $\lambda6716/\lambda6731$ ratio. Although the [S\,{\sc ii}] lines may be affected by narrow telluric absorption, independent diagnostics also favor substantially higher densities. CLASSY spectroscopy gives $n_{\rm e}({\rm [Ar\,IV]})\simeq1.1\times10^{3}~{\rm cm}^{-3}$, $n_{\rm e}({\rm [Fe\,III]})\simeq2.4\times10^{3}~{\rm cm}^{-3}$, and $n_{\rm e}({\rm C\,III]})\simeq4.3\times10^{3}~{\rm cm}^{-3}$ \citep{Mingozzi2022}. Spatially resolved measurements of the central super star clusters similarly give $n_{\rm e}\simeq250$--$320~{\rm cm}^{-3}$ from [O\,{\sc ii}] and [S\,{\sc ii}], but $n_{\rm e}\simeq2.4\times10^{3}~{\rm cm}^{-3}$ from [Ar\,{\sc iv}] \citep{Peng2026}. Thus, even the conservative optical--far-infrared [O\,{\sc iii}] upper limit lies below the electron densities indicated by several independent diagnostics.

As an additional check on the relative flux calibration over the wide MUSE wavelength range for SBS~0335--052E, the [S\,{\sc iii}] $\lambda6312/\lambda9069$ ratio gives $T_{\rm e}({\rm S^{++}})\simeq1.7\times10^{4}$~K. Adopting this lower temperature instead of $T_{\rm e}({\rm O^{++}})$ shifts the predicted [O\,{\sc iii}] 88~$\micron/\lambda5007$ ratio only modestly and does not remove the preference for a low effective electron density.

Haro~11 shows a similarly strong discrepancy among the different diagnostics. The nominal optical--far-infrared [O\,{\sc iii}] ratio favors $n_{\rm e}<1~{\rm cm}^{-3}$, while the conservative treatment of the observational uncertainties allows densities up to $n_{\rm e}\lesssim10~{\rm cm}^{-3}$. In comparison, the [O\,{\sc ii}] $\lambda3729/\lambda3726$ ratio gives $n_{\rm e}=240^{+110}_{-100}~{\rm cm}^{-3}$, and the [S\,{\sc ii}] $\lambda6716/\lambda6731$ ratio gives $870^{+90}_{-80}~{\rm cm}^{-3}$. We regard the [O\,{\sc ii}]-based density as the more robust low-ionization optical constraint because the [S\,{\sc ii}] lines may be affected by telluric absorption. At still higher densities, CLASSY spectroscopy of knot~C gives $n_{\rm e}({\rm [Cl\,III]})\sim4\times10^{3}~{\rm cm}^{-3}$ \citep{Berg2022,Mingozzi2022}. No useful ultraviolet density constraint is available for Haro~11 because significant UV emission lines other than Ly$\alpha$ are not detected. Thus, the effective electron density implied by the optical--far-infrared [O\,{\sc iii}] emission remains substantially below the densities indicated by the independent optical diagnostics even when the conservative [O\,{\sc iii}] upper limit is adopted.

By contrast, POX~186 shows no significant discrepancy between the optical--far-infrared [O\,{\sc iii}] density and the available optical and ultraviolet constraints. We obtain $n_{\rm e}=140^{+70}_{-60}~{\rm cm}^{-3}$ from the optical--far-infrared [O\,{\sc iii}] emission, consistent with $n_{\rm e}=240^{+170}_{-140}~{\rm cm}^{-3}$ from [S\,{\sc ii}]. The [C\,{\sc iii}] $\lambda1907$/C\,{\sc iii}] $\lambda1909$ ratio is also consistent with the low-density limit, corresponding to $n_{\rm e}({\rm C\,III]})=1^{+2300}_{-1}~{\rm cm}^{-3}$ \citep{Rogers2023}. The [N\,{\sc iv}] $\lambda1483$/N\,{\sc iv}] $\lambda1486$ ratio does not provide a quantitative density measurement because the $\lambda1486$ line is not detected, but its lower limit is likewise consistent with the low-density regime. The available diagnostics therefore provide no clear evidence for a substantially denser component in POX~186.

For I~Zw~18, the optical--far-infrared [O\,{\sc iii}] and low-ionization optical diagnostics are also mutually consistent, but higher-ionization diagnostics reveal a much broader density range. The [O\,{\sc iii}] diagnostic gives $n_{\rm e}=3^{+59}_{-2}~{\rm cm}^{-3}$, while [S\,{\sc ii}] gives $110^{+160}_{-80}~{\rm cm}^{-3}$. CLASSY measurements similarly give $n_{\rm e}({\rm [O\,II]})\sim10^{2}~{\rm cm}^{-3}$, whereas [C\,{\sc iii}] $\lambda1907$/C\,{\sc iii}] $\lambda1909$ and [Ar\,{\sc iv}] indicate densities of order $10^{4}~{\rm cm}^{-3}$ \citep{Mingozzi2022}. Spatially resolved KCWI spectroscopy likewise finds higher densities from [Ar\,{\sc iv}] than from [O\,{\sc ii}] \citep{RickardsVaught2025}. Thus, agreement between the galaxy-integrated optical--far-infrared [O\,{\sc iii}] emission and the low-ionization optical diagnostics does not imply that the ionized gas can be characterized by a single electron density.

Taken together, these comparisons demonstrate that the electron density inferred for a given galaxy can depend strongly on the adopted emission-line diagnostic. In SBS~0335--052E and Haro~11, the effective densities implied by the optical--far-infrared [O\,{\sc iii}] emission are substantially lower than those indicated by independent optical and ultraviolet diagnostics, whereas POX~186 shows no comparable discrepancy. I~Zw~18 further demonstrates that consistency between the [O\,{\sc iii}] and low-ionization diagnostics can coexist with much higher densities inferred from higher-ionization tracers. Such diagnostic-dependent electron densities may reflect differences among ionization zones and/or the different density regimes preferentially weighted by individual line ratios. A systematic hierarchy of inferred densities can also arise from a broad underlying density distribution without requiring sharply separated physical components \citep{MendezDelgado2026}. We discuss the physical implications of these results in section~\ref{sec:4.1}.

\begin{table}[t]
\tbl{Electron densities inferred from optical, ultraviolet, and optical--far-infrared diagnostics.}{%
\centering
\begin{tabular}{llc}
\hline
Object & Diagnostic & $n_{\rm e}$ [cm$^{-3}$] \\
\hline\hline
SBS 0335--052E
& [O\,{\sc iii}] 88~$\micron/\lambda5007$ & $\lesssim40$ \\
& [S\,{\sc ii}] $\lambda6716/\lambda6731$ & $290^{+360}_{-210}$ \\
& [Ar\,{\sc iv}] $\lambda4711/\lambda4740$ & $\simeq1.1\times10^{3}$ \\
& [Fe\,{\sc iii}] & $\simeq2.4\times10^{3}$ \\
& [C\,{\sc iii}] $\lambda1907$/C\,{\sc iii}] $\lambda1909$ & $\simeq4.3\times10^{3}$ \\
\hline
POX 186
& [O\,{\sc iii}] 88~$\micron/\lambda5007$ & $140^{+70}_{-60}$ \\
& [S\,{\sc ii}] $\lambda6716/\lambda6731$ & $240^{+170}_{-140}$ \\
& [C\,{\sc iii}] $\lambda1907$/C\,{\sc iii}] $\lambda1909$ & $1^{+2300}_{-1}$ \\
\hline
Haro 11
& [O\,{\sc iii}] 88~$\micron/\lambda5007$ & $\lesssim10$ \\
& [O\,{\sc ii}] $\lambda3729/\lambda3726$ & $240^{+110}_{-100}$ \\
& [S\,{\sc ii}] $\lambda6716/\lambda6731$ & $870^{+90}_{-80}$ \\
& [Cl\,{\sc iii}] $\lambda5518/\lambda5538$ & $\sim4\times10^{3}$ \\
\hline
I Zw 18
& [O\,{\sc iii}] 88~$\micron/\lambda5007$ & $3^{+59}_{-2}$ \\
& [S\,{\sc ii}] $\lambda6716/\lambda6731$ & $110^{+160}_{-80}$ \\
& [O\,{\sc ii}] $\lambda3729/\lambda3726$ & $\sim10^{2}$ \\
& [Ar\,{\sc iv}] $\lambda4711/\lambda4740$ & $\sim10^{4}$ \\
& [C\,{\sc iii}] $\lambda1907$/C\,{\sc iii}] $\lambda1909$ & $\sim10^{4}$ \\
\hline
\end{tabular}
}
\begin{tabnote}
\begin{minipage}[t]{0.9\linewidth}
\raggedright
{\bf Notes:}
The [O\,{\sc iii}] 88~$\micron/\lambda5007$, [S\,{\sc ii}], and [O\,{\sc ii}] densities from the present analysis are based on galaxy-integrated or nearly galaxy-integrated measurements.
For SBS~0335--052E and Haro~11, the nominal [O\,{\sc iii}] line ratios favor $n_{\rm e}<1~{\rm cm}^{-3}$; the values listed in the table are conservative upper limits obtained by adopting the lower $1\sigma$ limits of both [O\,{\sc iii}] ratios, as described in section~\ref{subsec:results3.1}.
For SBS~0335--052E, the [Ar\,{\sc iv}], [Fe\,{\sc iii}], and C\,{\sc iii}] densities are adopted from \citet{Mingozzi2022}.
For POX~186, the C\,{\sc iii}] density is adopted from \citet{Rogers2023}.
For Haro~11, the [Cl\,{\sc iii}] density is adopted from \citet{Mingozzi2022}.
For I~Zw~18, the [O\,{\sc ii}], [Ar\,{\sc iv}], and C\,{\sc iii}] densities are adopted from \citet{Mingozzi2022}.
These literature measurements generally probe different apertures and are therefore shown as complementary constraints rather than direct aperture-matched comparisons.
The [O\,{\sc iii}] 88~$\micron/\lambda5007$ values represent effective electron densities inferred under the one-zone assumption.
\end{minipage}
\end{tabnote}
\label{tab:tene_comparison}
\end{table}

\subsection{Two-zone Models for SBS~0335--052E and Haro~11}
\label{subsec:results3.3}

We further examined whether unresolved ionized-gas structure can reproduce the three observed [O\,{\sc iii}] lines in SBS~0335--052E and Haro~11. As shown in sections~\ref{subsec:results3.1} and \ref{subsec:results3.2}, their nominal [O\,{\sc iii}] line ratios favor the low-electron-density limit, corresponding to effective electron densities of $n_{\rm e}<1~{\rm cm}^{-3}$, while a conservative treatment of the observational uncertainties allows values up to approximately $40~{\rm cm}^{-3}$ and $10~{\rm cm}^{-3}$ for SBS~0335--052E and Haro~11, respectively. These effective densities remain substantially below those inferred from independent optical and ultraviolet diagnostics. Motivated by this discrepancy, we tested whether the observed emission can be reproduced by combining a relatively dense component with diffuse, low-density gas. Following \citet{Usui2025}, we modeled the total [O\,{\sc iii}] $\lambda4363$, $\lambda5007$, and 88~$\micron$ emission as the sum of two components with different electron temperatures, electron densities, and effective emitting volumes.

For a given [O\,{\sc iii}] transition $i$, the total line luminosity is expressed as
\begin{equation}
L_i
=
j_i(T_{{\rm e},1},n_{{\rm e},1})V_1
+
j_i(T_{{\rm e},2},n_{{\rm e},2})V_2,
\end{equation}
where $j_i(T_{\rm e},n_{\rm e})$ is the volume emissivity of transition $i$. For a collisionally excited [O\,{\sc iii}] line, we calculate it from the \texttt{PyNeb} emissivity coefficient $\epsilon_i(T_{\rm e},n_{\rm e})$ as
\begin{equation}
j_i(T_{\rm e},n_{\rm e})
=
\epsilon_i(T_{\rm e},n_{\rm e})\,
n_{\rm e}\,
n({\rm O^{++}}).
\end{equation}
Assuming the same O$^{++}$/H abundance for the two components within each galaxy, the predicted ratio of two transitions $i$ and $k$ is therefore
\begin{equation}
\frac{L_i}{L_k}
=
\frac{
j_i(T_{{\rm e},1},n_{{\rm e},1})(V_1/V_2)
+
j_i(T_{{\rm e},2},n_{{\rm e},2})
}{
j_k(T_{{\rm e},1},n_{{\rm e},1})(V_1/V_2)
+
j_k(T_{{\rm e},2},n_{{\rm e},2})
}.
\end{equation}

\begin{figure*}[!htbp]
\centering
\includegraphics[width=1.03\textwidth]{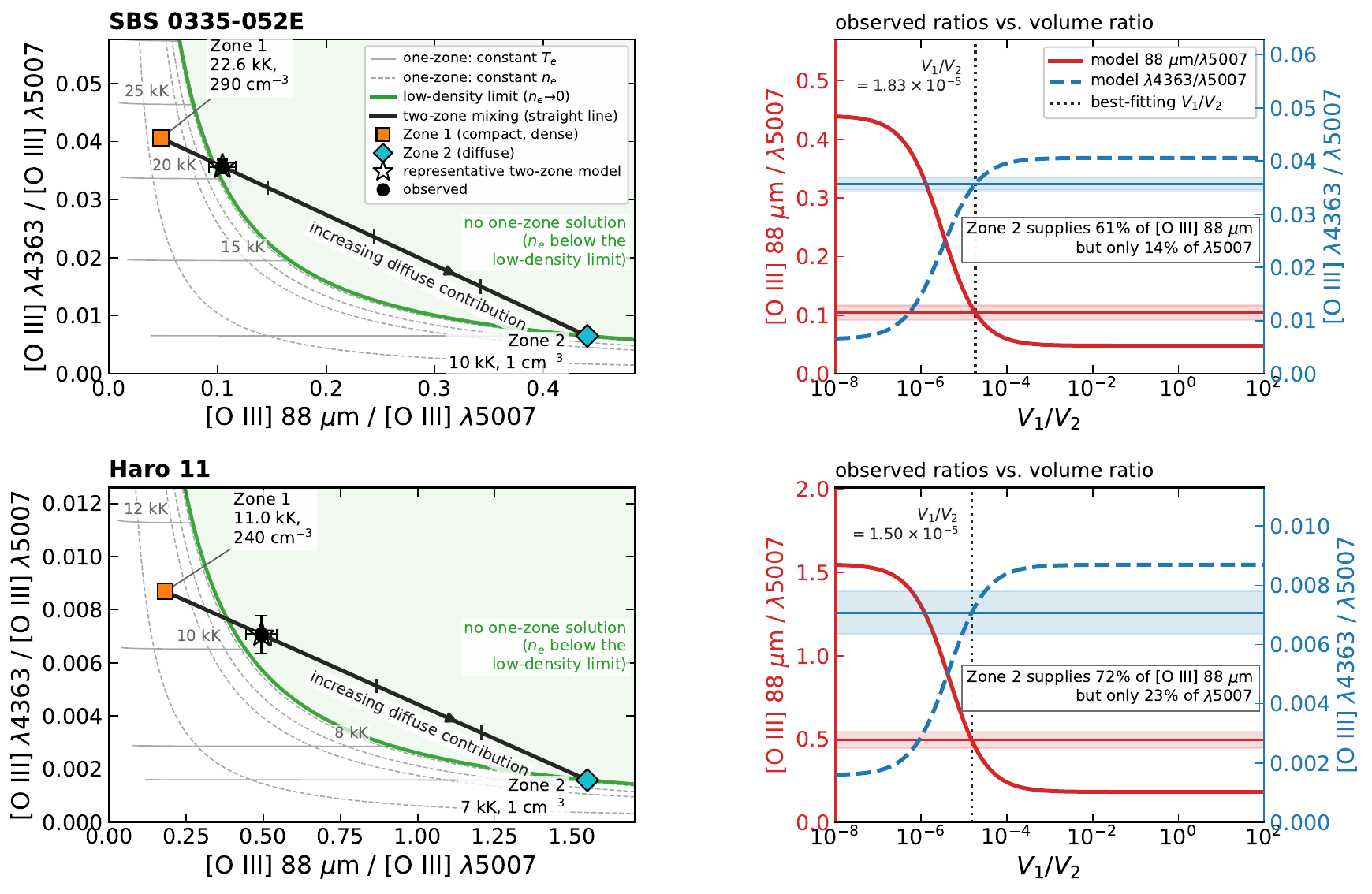}
\caption{
Two-zone models for SBS~0335--052E (top) and Haro~11 (bottom). In each row, the left panel shows the observed [O\,{\sc iii}] $\lambda4363$/[O\,{\sc iii}] $\lambda5007$ and [O\,{\sc iii}] 88~$\micron$/[O\,{\sc iii}] $\lambda5007$ ratios together with the mixing sequence connecting the adopted Zone~1 and Zone~2 conditions. The right panels show the predicted [O\,{\sc iii}] 88~$\micron$/[O\,{\sc iii}] $\lambda5007$ and [O\,{\sc iii}] $\lambda4363$/[O\,{\sc iii}] $\lambda5007$ ratios, respectively, as functions of the effective emitting-volume ratio $V_1/V_2$. The horizontal solid lines and shaded regions indicate the observed ratios and their $1\sigma$ uncertainties, respectively, while the vertical dashed lines mark the best-fitting effective emitting-volume ratios. For the [O\,{\sc iii}] 88~$\micron$/[O\,{\sc iii}] $\lambda5007$ ratio, the observational uncertainty includes the same additional 10\% systematic uncertainty adopted in figures~\ref{fig:OIIIdiagram} and \ref{fig:Tene} to account for possible relative flux-calibration differences between the optical and far-infrared measurements. The adopted two-zone solutions reproduce both observed [O\,{\sc iii}] line ratios simultaneously. The inferred temperatures, densities, and effective emitting-volume ratios should be regarded as representative rather than unique because the available three [O\,{\sc iii}] lines do not fully constrain all two-zone model parameters.
%
}
\label{fig:twozone}
\end{figure*}

To construct physically motivated solutions, we anchored the Zone~1 density to the optical density constraint for each galaxy: $n_{{\rm e},1}=290~{\rm cm}^{-3}$ for SBS~0335--052E, motivated by the [S\,{\sc ii}] measurement, and $n_{{\rm e},1}=240~{\rm cm}^{-3}$ for Haro~11, based on the more robust [O\,{\sc ii}] measurement. We then searched for temperatures, diffuse-component densities, and volume ratios that reproduce both the observed [O\,{\sc iii}] $\lambda4363/\lambda5007$ and 88~$\micron/\lambda5007$ ratios. Because the parameters are not uniquely constrained, the solutions shown below are intended as representative examples rather than unique decompositions of the ionized gas. Figure~\ref{fig:twozone} shows representative solutions for SBS~0335--052E in the top panels and Haro~11 in the bottom panels. The left panels show the mixing sequences in the [O\,{\sc iii}] diagnostic plane, while the right panels show the two line ratios as functions of $V_1/V_2$.

For SBS~0335--052E, a representative solution combines $(T_{{\rm e},1},n_{{\rm e},1})=(22600~{\rm K},290~{\rm cm}^{-3})$ with a diffuse component having $(T_{{\rm e},2},n_{{\rm e},2})=(10000~{\rm K},1~{\rm cm}^{-3})$, with $V_1/V_2\simeq1.8\times10^{-5}$. The diffuse component contributes approximately 61\% of the total [O\,{\sc iii}] 88~$\micron$ luminosity but only approximately 14\% of the [O\,{\sc iii}] $\lambda5007$ luminosity.

For Haro~11, a representative solution combines $(T_{{\rm e},1},n_{{\rm e},1})=(11000~{\rm K},240~{\rm cm}^{-3})$ with $(T_{{\rm e},2},n_{{\rm e},2})=(7000~{\rm K},1~{\rm cm}^{-3})$, with $V_1/V_2\simeq1.5\times10^{-5}$. The diffuse component contributes approximately 72\% of the [O\,{\sc iii}] 88~$\micron$ luminosity but only approximately 23\% of the [O\,{\sc iii}] $\lambda5007$ luminosity. Thus, in both galaxies, diffuse low-density gas can dominate the far-infrared [O\,{\sc iii}] emission while contributing only a minority of the optical [O\,{\sc iii}] emission.

As a consistency check on the physical scale of the diffuse component, we estimated its effective emitting volume from the absolute [O\,{\sc iii}] 88~$\micron$ luminosity. We used the fraction of the total 88~$\micron$ luminosity contributed by Zone~2 in each representative solution and evaluated

\begin{equation}
L_{88,2}
=
\epsilon_{88}(T_{{\rm e},2},n_{{\rm e},2})
n_{{\rm e},2}n({\rm O^{++}})V_{{\rm emit},2}.
\end{equation}

where $\epsilon_{88}$ is the [O\,{\sc iii}] 88~$\micron$ emissivity coefficient per $n_{\rm e}n({\rm O^{++}})$, calculated for the adopted Zone~2 conditions. The O$^{++}$ number density was estimated from the adopted gas-phase oxygen abundance and O$^{++}$/O ionic fraction, assuming $n_{\rm e}\simeq1.08n_{\rm H}$ to account approximately for the contribution of helium to the free-electron density. For SBS~0335--052E, we adopted $12+\log({\rm O/H})=7.25$ and ${\rm O^{++}/O}\simeq0.88$, consistent with the ionic abundances measured in different regions of the galaxy by \citet{Papaderos2006}. For Haro~11, we adopted $12+\log({\rm O/H})=8.36$ and a representative ${\rm O^{++}/O}\simeq0.60$, based on the ionic abundances reported for its bright H\,{\sc ii} regions by \citet{Guseva2012}. For the representative $n_{{\rm e},2}=1~{\rm cm}^{-3}$ solutions, the resulting emitting volumes correspond to equivalent spherical radii of approximately 1.7~kpc for SBS~0335--052E and 3.5~kpc for Haro~11. These characteristic scales are comparable to the extended ionized-gas structures observed with MUSE: ionized filaments extend over several kiloparsecs in SBS~0335--052E \citep{Herenz2017,Herenz2023}, while Haro~11 exhibits a kiloparsec-scale ionized-gas halo containing bubbles, filaments, arcs, and channels \citep{Menacho2019,Sirressi2022}. Thus, the representative solutions do not require obviously implausible emitting volumes.

These equivalent radii refer to the line-emitting volume and implicitly correspond to a volume filling factor of unity. More generally, the [O\,{\sc iii}] luminosity constrains the product $fV_{\rm geom}$, where $f$ is the volume filling factor and $V_{\rm geom}$ is the geometric volume occupied by the diffuse component. In the low-density regime, where the line luminosity scales approximately as $L_{88}\propto n_{\rm e}^{2}fV_{\rm geom}$, the corresponding characteristic geometric radius scales as $R_{\rm geom}\propto n_{{\rm e},2}^{-2/3}f^{-1/3}$. The absolute spatial scale is consequently degenerate with both the assumed diffuse-gas density and filling factor, and the radii above should be regarded as order-of-magnitude consistency checks rather than direct size measurements.

The two-zone parameters are not uniquely constrained, particularly because the density of the diffuse component is unknown. The inferred effective volume and mass ratios therefore depend strongly on the assumed $n_{{\rm e},2}$ and on the filling factors of the two components and should be regarded as illustrative. The adopted value of $n_{{\rm e},2}=1~{\rm cm}^{-3}$ in the representative solutions is motivated by the nominal [O\,{\sc iii}] line ratios but is not uniquely required by the data. By contrast, the relative line-luminosity contributions are much less sensitive to this assumption: over the tested range $n_{{\rm e},2}=0.1$--$40~{\rm cm}^{-3}$, the diffuse component contributes approximately 60--75\% of the [O\,{\sc iii}] 88~$\micron$ emission while contributing only approximately 10--25\% of the [O\,{\sc iii}] $\lambda5007$ emission. We therefore regard this differential contribution as the more robust result of the two-zone modeling.

The two components need not correspond to physically discrete gas zones, but may instead provide a simple representation of a broader unresolved distribution of gas densities and temperatures in which different emission-line diagnostics preferentially weight different density regimes \citep{MendezDelgado2026}. Other processes affecting diffuse ionized gas, such as shocks or spectral hardening of the ionizing radiation field, may also contribute, and the present data do not distinguish among these possibilities. Direct measurements of [O\,{\sc iii}] 52~$\micron$/88~$\micron$ would provide an important independent constraint on the density of the far-infrared-emitting gas. The present two-zone models are constrained only by the three [O\,{\sc iii}] lines and are therefore not intended to reproduce the full emission-line spectrum. Testing whether the same physical components can simultaneously reproduce other optical and ultraviolet diagnostics would require more detailed photoionization modeling and is beyond the scope of this study. Moreover, the available IFU data do not allow us to identify a single spatial region with fixed $T_{\rm e}$ and $n_{\rm e}$ that dominates all three [O\,{\sc iii}] lines, because the 88~$\micron$ measurements are galaxy-integrated and [O\,{\sc iii}] $\lambda4363$ generally lacks sufficient spatially resolved signal-to-noise.

\section{Discussion}
\label{sec:discussion4}

\subsection{Diversity of [O\,{\sc iii}]-emitting Gas Structures}
\label{sec:4.1}

Nearby low-metallicity galaxies show substantial diversity in the consistency of their optical and far-infrared ionized-gas diagnostics. For SBS~0335--052E and Haro~11, the nominal three-line [O\,{\sc iii}] ratios drive the one-zone solutions toward the low-electron-density limit, corresponding to effective electron densities of $n_{\rm e}<1~{\rm cm}^{-3}$. Even when the observational uncertainties are treated conservatively, values are limited to approximately $n_{\rm e}\lesssim40~{\rm cm}^{-3}$ and $\lesssim10~{\rm cm}^{-3}$ for SBS~0335--052E and Haro~11, respectively. These effective densities remain substantially below those indicated by independent optical and ultraviolet diagnostics. By contrast, POX~186 and I~Zw~18 show no significant discrepancy between the optical--far-infrared [O\,{\sc iii}] diagnostic and the available low-ionization optical diagnostics within the uncertainties. This contrasts with the broad agreement previously reported between densities inferred from optical doublets and from the far-infrared [O\,{\sc iii}] 52~$\micron$/88~$\micron$ ratio for most nearby galaxies examined \citep{Harikane2025}.

A key advantage of our analysis is the reduction of aperture mismatches. The nearby comparison sample of \citet{Harikane2025} relies largely on the ISO/LWS compilation of \citet{Brauher2008}, for which the far-infrared measurements are not always galaxy-integrated. In several cases, the optical extent exceeds the $\sim75''$ ISO/LWS beam, and the fluxes were derived using point-source calibration without an extended-source correction. By contrast, for our primary sample we combine galaxy-integrated \textit{Herschel}/PACS measurements with optical IFU data or integrated spectroscopy chosen to provide comparable spatial coverage, thereby reducing one of the principal systematic uncertainties in optical--far-infrared comparisons.

As shown in section~\ref{subsec:results3.2}, however, aperture matching alone does not guarantee that different emission-line diagnostics yield the same electron density. The inferred density can depend strongly on the adopted diagnostic, with low-ionization optical, higher-ionization optical and ultraviolet, and optical--far-infrared [O\,{\sc iii}] ratios sampling markedly different density regimes in some galaxies. Such differences may arise from ionization-dependent density structure and/or because individual diagnostics preferentially weight different parts of a broad underlying density distribution \citep{MendezDelgado2026}. We therefore examine below whether the observed internal structures of these galaxies provide a physical basis for these diagnostic differences.

\subsubsection{Evidence for Complex Ionized-gas Structure in Nearby Galaxies}

Haro~11 and SBS~0335--052E both show substantial evidence for complex ionized-gas structure. In Haro~11, spatially resolved MUSE observations reveal strong variations in $n_{\rm e}$, $T_{\rm e}$, dust attenuation, and ionization conditions across the three star-forming knots and surrounding ionized gas, together with superbubbles, shells, filaments, ionized cones, and outflows \citep{Menacho2019,Menacho2021}. Knot~C occupies a particularly low-density cavity \citep{Menacho2021}. SBS~0335--052E likewise exhibits a compact, heavily obscured source associated with SSC~1, surrounded by less obscured high-ionization emission in JWST/MIRI observations \citep{Mingozzi2025}. The hard ionizing source may be an obscured intermediate-mass black hole or an $\eta$~Carinae-like stellar eruption \citep{Mingozzi2025,Hatano2026,Peng2026}. Together with the broad range of electron densities inferred from independent optical and ultraviolet diagnostics (section~\ref{subsec:results3.2}), these observations provide a natural physical basis for the low effective electron densities inferred from the optical--far-infrared [O\,{\sc iii}] emission in the two galaxies. The representative two-zone models in section~\ref{subsec:results3.3} illustrate one possible realization of such unresolved structure.

Differential dust attenuation may provide an additional contribution, particularly in SBS~0335--052E. Its infrared spectral energy distribution indicates heavily obscured star formation \citep{Thuan1999}, and emission from strongly obscured H\,{\sc ii} regions could contribute to [O\,{\sc iii}] 88~$\micron$ while their [O\,{\sc iii}] $\lambda5007$ emission is strongly attenuated. Such emission would not be recovered by the Balmer-decrement correction if the corresponding optical component were effectively hidden. Haro~11 also contains warm dust and obscured star formation, although detailed modeling of its \textit{Spitzer} and \textit{Herschel} observations found that the [O\,{\sc iii}] 88~$\micron$ emission is not fully reproduced by the main compact H\,{\sc ii}-region component and suggested an additional low-density, high-ionization medium \citep{Cormier2012}. Thus, differential obscuration may contribute to the observed optical--far-infrared [O\,{\sc iii}] ratios, but unresolved density and ionization structure may also play an important role.

POX~186 and I~Zw~18 provide useful contrasts. POX~186 is extremely compact, with a maximum physical extent of only $\sim300$~pc and most of its star formation concentrated in a central star cluster with a diameter of $\sim10$--15~pc \citep{Corbin2002}. Its optical--far-infrared [O\,{\sc iii}] density is consistent with the available optical and ultraviolet density constraints within the uncertainties. I~Zw~18 also shows consistency between the integrated [O\,{\sc iii}] and low-ionization optical diagnostics, while higher-ionization optical and ultraviolet diagnostics indicate substantially denser gas \citep{RickardsVaught2025,Hunt2025a,Hunt2025b}. This illustrates how different line ratios can preferentially weight different parts of the underlying density distribution, particularly because [O\,{\sc iii}] 88~$\micron$ becomes progressively collisionally suppressed at densities above its critical density.

Taken together, these galaxies suggest that the diversity seen in integrated ionized-gas diagnostics can arise naturally from unresolved variations in density, temperature, attenuation, and ionization conditions. The representative two-zone solutions provide one simple description of this complexity, while a broader continuous distribution of gas properties is also possible \citep{MendezDelgado2026}. The $T_{\rm e}$ and $n_{\rm e}$ inferred from integrated one-zone analyses are therefore best interpreted as effective quantities that characterize the line-emitting gas weighted by each diagnostic.

\subsection{Implications for High-redshift Galaxies}

Figure~\ref{fig:OIIIdiagram} enables a direct comparison between our nearby galaxies and high-redshift galaxies in the same optical--far-infrared [O\,{\sc iii}] diagnostic plane. High-redshift galaxies span a broad range of positions relative to the one-zone model grid. Our nearby sample shows that similarly low effective [O\,{\sc iii}] electron densities and discrepancies with independent density diagnostics can also occur in the local Universe. In particular, SBS~0335--052E and Haro~11 lie near or slightly beyond the low-electron-density boundary of the model grid and show substantially lower effective optical--far-infrared [O\,{\sc iii}] densities than those inferred from other diagnostics. Thus, such behavior is not restricted to high-redshift galaxies.

Their physical origin, however, may differ among galaxies. In COS-2987 at $z=6.81$, the optical and far-infrared [O\,{\sc iii}] emission has been interpreted in terms of components with different temperatures and densities \citep{Usui2025}, while the lack of detected dust continuum disfavors strong differential obscuration as the primary explanation \citep{Witstok2022}. By contrast, MACS0416-Y1 at $z=8.31$ is consistent with a one-zone model in the optical--far-infrared [O\,{\sc iii}] plane, while its [O\,{\sc ii}]-based density is inconsistent with the effective density implied by the [O\,{\sc iii}] emission \citep{Tamura2019,Bakx2020,Tamura2023,Bakx2025,Takechi2026}. Such discrepancies need not imply sharply separated components because different diagnostics may preferentially weight different parts of a broad density distribution \citep{MendezDelgado2026}.

Morphological complexity and AGN activity may also contribute to the observed diversity \citep{Harikane2025}. COS-2987 and COS-3018 are resolved into multiple components \citep{Usui2025,Mawatari2026,Scholtz2025}, while J0217-0208 may host an AGN \citep{Harikane2025}. However, the present samples are too small to establish a direct connection between morphology, AGN activity, and the behavior of the optical--far-infrared density diagnostics. Spatially resolved observations will be required to determine which physical structures dominate the emission measured in integrated spectra.

The combined nearby and high-redshift comparison therefore suggests that discrepancies among optical and far-infrared ionized-gas diagnostics can arise from unresolved variations in gas density, temperature, attenuation, and ionization conditions and are not restricted to high redshift. Distinguishing among these effects will require larger matched-aperture samples and ultimately spatially resolved measurements of both optical and far-infrared ionized-gas tracers.

\section{Conclusions}
\label{sec:conclusions5}

We investigated galaxy-integrated optical and far-infrared [O\,{\sc iii}] emission in five nearby metal-poor dwarf galaxies---HS~1222+3741, SBS~0335--052E, POX~186, Haro~11, and I~Zw~18---to examine whether their integrated ionized-gas emission can be represented by a single electron temperature and density. We combined [O\,{\sc iii}] $\lambda4363$ and $\lambda5007$ measurements with \textit{Herschel}/PACS [O\,{\sc iii}] 88~$\micron$ data, using Seimei/KOOLS-IFU observations and archival or published spectroscopy with comparable spatial coverage. The KOOLS-IFU observations provide, to our knowledge, the first optical integral-field spectroscopy of HS~1222+3741. Because [O\,{\sc iii}] $\lambda4363$ is not detected in HS~1222+3741, the quantitative optical--far-infrared analysis is based on the remaining four galaxies. Our main conclusions are as follows:

\begin{enumerate}

\item
\textbf{Diversity in optical--far-infrared [O\,{\sc iii}] diagnostics.}
SBS~0335--052E and Haro~11 lie near or slightly beyond the low-electron-density boundary of the one-zone [O\,{\sc iii}] diagnostic. Their nominal line ratios favor effective electron densities of $n_{\rm e}<1~{\rm cm}^{-3}$, while a conservative treatment of the observational uncertainties allows values up to approximately $n_{\rm e}\lesssim40~{\rm cm}^{-3}$ and $\lesssim10~{\rm cm}^{-3}$, respectively. By contrast, POX~186 and I~Zw~18 are consistent with the one-zone [O\,{\sc iii}] diagnostic within the uncertainties. The [O\,{\sc iii}] $\lambda4363$ upper limit for HS~1222+3741 does not provide a useful quantitative constraint on the one-zone interpretation.

\item
\textbf{Strong diagnostic dependence of the inferred electron density.}
Even when the conservative upper limits are adopted, the effective optical--far-infrared [O\,{\sc iii}] densities in SBS~0335--052E and Haro~11 remain substantially below those inferred from independent optical density diagnostics. Additional optical and ultraviolet diagnostics indicate still higher densities in these galaxies. By contrast, POX~186 shows no significant discrepancy among the available constraints, while I~Zw~18 demonstrates that agreement between the integrated [O\,{\sc iii}] and low-ionization optical diagnostics can coexist with much higher densities inferred from higher-ionization tracers. These results show that the electron density inferred for a galaxy can depend strongly on the adopted diagnostic, even when aperture mismatches are minimized.

\item
\textbf{Unresolved ionized-gas structure as a natural explanation.}
Representative two-zone models reproduce the observed [O\,{\sc iii}] $\lambda4363$, $\lambda5007$, and 88~$\micron$ emission in SBS~0335--052E and Haro~11. For SBS~0335--052E, a representative solution combines $(T_{\rm e},n_{\rm e})=(22600~{\rm K},290~{\rm cm}^{-3})$ and $(10000~{\rm K},1~{\rm cm}^{-3})$, while for Haro~11 it combines $(11000~{\rm K},240~{\rm cm}^{-3})$ and $(7000~{\rm K},1~{\rm cm}^{-3})$. In these solutions, the low-density component contributes approximately 61\% and 72\% of the [O\,{\sc iii}] 88~$\micron$ luminosity in SBS~0335--052E and Haro~11, respectively, but only approximately 14\% and 23\% of the [O\,{\sc iii}] $\lambda5007$ luminosity. The precise decomposition is not unique: a broader distribution of density and temperature, and in some cases differential dust attenuation, may produce similar optical--far-infrared behavior. Thus, the two-zone models should be regarded as representative descriptions of unresolved ionized-gas structure rather than unique physical decompositions.

\item
\textbf{Implications for high-redshift galaxies.}
The diversity observed among these nearby metal-poor galaxies provides a local counterpart to the range of optical--far-infrared [O\,{\sc iii}] properties observed at high redshift. In particular, the low effective [O\,{\sc iii}] electron densities and discrepancies with independent density diagnostics found in SBS~0335--052E and Haro~11 demonstrate that such behavior is not restricted to the early Universe. Interpreting spatially integrated JWST and ALMA measurements therefore requires consideration of unresolved density, temperature, attenuation, and ionization structure rather than assigning a unique $T_{\rm e}$ or $n_{\rm e}$ to all of the ionized gas.

\end{enumerate}

Larger samples with matched optical and far-infrared apertures, together with multiple density-sensitive diagnostics spanning different density regimes, will be essential for determining how commonly integrated galaxy spectra exhibit diagnostic-dependent electron densities and for quantifying their impact on physical properties inferred from emission-line measurements.

\begin{ack}

We thank Cassandra Barlow-Hall, Dongsheng Sun, Mitsutaka Usui, Asahi Hamada, Yuruzu Terui, Takeshi Hashigaya, Nario Kuno, Shunsuke Honda, Yuri Nishimura, Yuichi Harikane, and Masami Ouchi for providing valuable comments on presentations by the first author.
We also thank Atsushi Yasuda, Hiroma Okubo, Nozomi Ishii, Saho Kawahara, and Ryota Ura for participating in the observations.
We thank Hideyuki Izumiura, Hiroyuki Maehara, Akito Tajitsu, Masaaki Otsuka, and Keisuke Isogai at Okayama Observatory, a branch of the National Astronomical Observatory of Japan, for their support during the observations.
T.H. was supported by the Leading Initiative for Excellent Young Researchers, MEXT, Japan (HJH02007), and by JSPS KAKENHI Grant Numbers 22H01258, 23K22529, and 25K00020. T.H. was also supported by the ALMA Japan Research Grant of NAOJ ALMA Project, NAOJ-ALMA-2025-28A. 
We acknowledge support from MEXT/JSPS KAKENHI Grant Numbers 22H04939 (M.H.), 23H00131 and 26H02069 (A.K.I.), 26K07139 (K.M.), 26K17200 (Y.S.), and 26H02061 (H.Y.). Y.N. acknowledges support from the Flatiron Research Fellowship. The Flatiron Institute is a division of the Simons Foundation. W.O. acknowledges support from JST SPRING, Grant Number JPMJSP2124. H.Y. also acknowledges support from the JST FOREST Program (Grant Number JPMJFR202Z).
\end{ack}

\bibliographystyle{apj}
\bibliography{ms}

\end{document}